\documentclass[paper]{JHEP}
\usepackage{amsmath}
\usepackage{epsfig}

\def\al{\alpha}
\def\as{\alpha_{\mbox{\scriptsize s}}}
\def\aef{\alpha_{\mbox{\scriptsize eff}}}
\def\daef{\delta \aef}
\def\qq{q\bar{q}}
\def\ee{e^+e^-}
\def\LQCD{\Lambda_{\mbox{\scriptsize QCD}}}
\def\MSbar{\overline{\mbox{MS}}}
\def\MSbar{\overline{\mbox{\scriptsize MS}}}
\def\GeV{\mathop{\rm Ge\!V}}
\def\xB{x_{\mbox{\scriptsize B}}}
\def\bq{\bar q}

\def\can{\tau}

\def\be{\beta}

\def\eps{\epsilon}
\def\de{\delta}

\def\om{\omega}
\def\gam{\gamma}
\def\lam{\lambda}

\def\mat{\mbox{\scriptsize mat}}
\def\res{\mbox{\scriptsize res}}

\def\cO#1{{\cal{O}}\left(#1\right)}
\def\half{\mbox{\small $\frac{1}{2}$}}

\def\VEV#1{\left\langle#1\right\rangle}

\def\PT{\mbox{\scriptsize PT}}
\def\NP{\mbox{\scriptsize NP}}

\def\conf{\delta}
\def\cp{\lambda^{\NP}}

\def\ka{\kappa}
\def\vka{\vec{\ka}}

\def\cF{{\cal{F}}}

\def\cM{{\cal{M}}}

\def\cR{{\cal{R}}}
\def\cP{{\cal {P}}}
\def\cI{{\cal {I}}}
\def\cC{{\cal {C}}}
\def\cS{{\cal {S}}}
\def\cA{{\cal{A}}}

\def\EET{{$\rm E_T \!E_T \!C$}}
 \newskip\humongous \humongous=0pt plus 1000pt minus 1000pt
   \newif\ifdtup

\def\la{\mathrel{\mathpalette\fun <}}
\def\ga{\mathrel{\mathpalette\fun >}}
\def\fun#1#2{\lower3.6pt\vbox{\baselineskip0pt\lineskip.9pt
  \ialign{$\mathsurround=0pt#1\hfil##\hfil$\crcr#2\crcr\sim\crcr}}}
\title{Azimuthal correlation in DIS\footnote{Research
    supported in part by the EU Fourth Framework Programme, `Training
    and Mobility of Researchers', Network `Quantum Chromodynamics and
    the Deep Structure of Elementary Particles', contract
    FMRX-CT98-0194 (DG12 - MIHT).}}

\author{
A.~Banfi, G.~Marchesini, G.~Smye\\
Dipartimento di Fisica, Universit{\`a} di Milano--Bicocca and \\
INFN, Sezione di Milano, Italy}

\abstract{We introduce the azimuthal correlation for the deep
  inelastic scattering process. We present the QCD prediction to the
  level of next-to-leading log resummation, matching to the fixed
  order prediction. We also estimate the leading non-perturbative
  power correction. The observable is compared with the energy-energy
  correlation in $\ee$ annihilation, on which it is modelled. The
  effects of the resummation and of the leading power correction are
  both quite large.  It would therefore be particularly instructive to
  study this observable experimentally.}

\keywords{QCD, Deep Inelastic Scattering, Jets, Nonperturbative Effects}

\preprint{
     Bicocca--FT--01/23\\
     hep-ph/0203150}

\begin{document}

\section{Introduction}
One of the first observables studied in QCD was the energy-energy
correlation (EEC) in $\ee$ annihilation \cite{EEC1,PP,EEC}. It is a function
of $\theta$, the polar angle between pairs of hadrons weighted by
their energy.
A similar correlation in DIS involves not a polar but an azimuthal
angle.  Here the observable is given by
\begin{equation}
\label{eq:H}
H(\chi)\! =\! \sin\chi\!\sum_{hh'}\frac{p_{th}p_{th'}}{Q^2}\,
\delta(\cos\chi\!+\!\cos\phi_{hh'})\,,
\quad \phi_{hh'}=\phi_h-\phi_{h'}\,,
\quad 0\le\chi\le\pi\,,
\end{equation}
where the sum runs over all outgoing hadron pairs with $p_{th}$ and
$\phi_{h}$ the transverse momentum and azimuthal angle in the Breit frame.
We take $-\pi\!<\!\phi_{hh'}\!\le\!\pi$.
The distribution for the observable $H(\chi)$ is dominated by events
with large $p_{t}$ jets: indeed we wish to exclude events with $p_t$
much less than $Q$ in order that we can perform the calculation based
on a dijet type configuration. On the other hand, to simplify the QCD
analysis and consider the incoming $\gam$ or $Z_0$ as pointlike, one
should limit the distribution to events in which all jet $p_{t}$ are
not much larger than $Q$. A convenient\footnote{This method to set a
limit on the large transverse momenta should contain small
hadronization corrections, see \cite{DMW,3ee,disko}.} way to set these
conditions is obtained by limiting the dijet resolution variable $y_2$
defined by the $k_t$ jet finding algorithm \cite{DISkt} as follows:
\begin{equation}
\label{eq:y2-lim}
y_-<y_2<y_+ \>.
\end{equation}
Here the lower bound $y_-$ selects large $p_t$ events, while the upper
bound $y_+$ allows us to consider the incoming vector boson as
pointlike.

The azimuthal differential distribution for this observable is then
defined, at fixed $\xB$ and $Q^2$, by
\begin{equation}
\label{eq:cH}
\frac{d\sigma(\chi,y_\pm)}{d\chi\,d\xB dQ^2} =
\sum_n\int\frac{d\sigma_n}{d\xB dQ^2}\,\Theta(y_+-y_2)\Theta(y_2-y_-)
\cdot H(\chi)\>,
\end{equation}
with $d\sigma_n/d\xB dQ^2$ the distribution for $n$ emitted hadrons in
the process under consideration.  One introduces the following
normalized azimuthal correlation (\EET)
\begin{equation}
\label{eq:Sigma}
\begin{split}
\frac{d\Sigma(\chi,y_\pm)}{d\chi}\>=\>
&\sigma^{-1}(y_\pm)\frac{d\sigma(\chi,y_\pm)}{d\chi\,d\xB dQ^2}\>,\\
\sigma(y_\pm)\>=\>
&\sum_n\int\frac{d\sigma_n}{d\xB dQ^2}\,\Theta(y_+-y_2)\Theta(y_2-y_-)
\left(\sum_{h=1}^n\frac{p_{th}}{Q}\right)^2.
\end{split}
\end{equation}
By taking into account the case $h=h'$ in \eqref{eq:H}, i.e.
$\chi=\pi$, this correlation is normalized to $1$. We will be
concerned with the back-to-back region of small $\chi=\pi-|\phi|$.

A standard method to start to understand the behaviour of this
correlation for small $\chi$ consists in trying to exponentiate the
leading logarithmic one-loop contribution given\footnote{Contributions
  from parton density functions in \EET\ enter beyond double
  logarithmic level.} by
\begin{equation} 
  \label{eq:Sig-1}
\Sigma(\chi,y_\pm)= 1-C_{T}\,\frac{\as}{4\pi}\,\ln^2\frac{1}{\chi^2}+\ldots\,, 
\quad C_T\!=\!2C_F\!+\!C_A\,,\quad \chi\ll1\>,
\end{equation}
where $C_T$ is the total colour charge of the three hard partons $\qq
g$ generating both the outgoing dijet and the incoming proton jet.
The exponentiation of this one-loop term gives a characteristic Sudakov
form factor leading to an azimuthal differential distribution
$d\Sigma/d\chi$ with a peak at small $\chi$ near
\begin{equation}
  \label{eq:peak}
  \chi \sim e^{-{\pi}/(2C_T\as)}\>.
\end{equation}
This is actually what happens for EEC in $\ee$ annihilation in the
back-to-back region (small $\chi=\pi-|\theta|$ with $\theta$ the polar
angle): the leading logarithmic one-loop contribution of
$\Sigma_{\ee}(\chi)$, is given by the same expression in
\eqref{eq:Sig-1} with $C_T=2C_F$, the charge of the $\qq$ pair in the
hard vertex. In this case the complete QCD analysis \cite{EEC1,PP,EEC}
and the data \cite{EECdata} confirm the presence in
$d\Sigma_{\ee}/d\chi$ of the Sudakov peak in \eqref{eq:peak}.

There are experimental indications that azimuthal distributions behave
differently. For instance, in $p\bar p$ processes, the differential
distribution in the azimuthal angle between two hadrons behaves
smoothly in the back-to-back region and shows no sign of a peak, as
reported for instance in \cite{EETdata}. We should then understand why
\EET\ and EEC behave differently.

In this paper we perform the QCD analysis of the DIS azimuthal
correlation \eqref{eq:Sigma} in the back-to-back region. The analysis
is similar to the one \cite{EEC1,PP,EEC} for EEC with a few major
differences:
\begin{itemize}
\item the distribution \eqref{eq:Sigma} is dominated by dijet events
  which originate from the $\qq g$ hard system ($3$-jet events
  including the beam).  Only recently has the QCD analysis been
  extended to distributions involving $3$ jets \cite{3ee,disko,hh} at
  the same accuracy as available in the study of $2$-jet CIS
  distributions \cite{PTstandard,TDIS};
\item the radiation off the incoming parton contributes not only to
  the observable $H(\chi)$, but also to the parton density evolution;
\item QCD resummation is performed in the $1$-dimensional impact
  parameter $b$ conjugate to the azimuthal angle. In the EEC case the
  variable is a polar angle and then one deals with the standard
  $2$-dimensional impact parameter.
\end{itemize}
As in other similar analyses, see \cite{disko,hh,TDIS}, due to QCD
coherence, the radiation factorizes giving rise to the evolved
standard parton density functions times the ``radiation factor'', a
collinear and infrared safe (CIS) quantity involving $3$-jet emission;

We work at the following accuracy: (i) double (DL) and single (SL)
logarithmic perturbative (PT) resummation (for the logarithm of the
distribution in the $b$-representation we resum all terms
$\as^n\ln^{n+1}\!bQ$ and $\as^n\ln^{n}\!bQ$); (ii) matching of
resummed with exact fixed order results (in this paper only to
one-loop order); (iii) leading non-perturbative (NP) corrections
coming from the fact that the QCD coupling runs into the infrared
region \cite{DMW,NPstandard}.

The paper is organized as follows.  In section 2 we discuss the QCD
process and the observable $H(\chi)$.  In section 3 we describe the
procedure for the resummation at SL level.  In section 4 we present
the result of the resummation and evaluate the leading power
correction.  In section 5 we present the numerical evaluation.  In
section 6 we the summarize and discuss our result.  Appendices contain
the technical details.

\section{QCD description \label{sec:QCD}}
We work in the Breit frame of the DIS process with $P$ and $q$ the
momenta of the incoming proton and the exchanged vector boson ($\gam$
or $Z_0$)
\begin{equation}
  \label{eq:Breit}
q = \frac{Q}{2}(0,0,0,2)\>, \qquad
P = \frac{Q}{2\xB}(1,0,0,-1) \>,\qquad
\xB = \frac{Q^2}{2(Pq)} \>.
\end{equation}
We first discuss the QCD hard elementary vertex and then the secondary
parton emission.

\subsection{Elementary hard process}
DIS dijet events originate from the elementary hard vertex 
\begin{equation}
  \label{eq:El-proc}
  q\,P_1\to P_2\,P_3\>,
\end{equation}
with $P_1$ along the $z$-axis. We take $P_2,P_3$ in the
$\{y,z\}$-plane, the ``event-plane'', see appendix \ref{App:El}. We
use the kinematical variables
\begin{equation}
  \label{eq:El-kin} 
\xi=\frac{(P_1 P_2)}{(P_1 q)} \>,\qquad 
x = \frac{Q^2}{2(P_1 q)}>\xB \>.
\end{equation}
The outgoing transverse momentum $P_t$ and the squared invariant
masses $Q^2_{ab}\!=\!2(P_aP_b)$ are given by
\begin{equation}
  \label{eq:Qab}
\begin{split}
P_t\!=\!Q\,\sqrt{\xi(1-\xi)\,\frac{1-x}{x}}\,,\qquad 
Q^2_{12}\!=\!\frac{\xi}{x}Q^2\,,\quad   
Q^2_{13}\!=\!\frac{1\!-\!\xi}{x}Q^2\,,\quad
Q^2_{23}\!=\!\frac{1\!-\!x}{x}Q^2\,.
\end{split}
\end{equation}
We distinguish $P_2$ from $P_3$ by assuming
\begin{equation}
  \label{eq:P2P3}
  (P_1P_2)\> < \>(P_1P_3)\>,
\end{equation}
which restricts us to the region $0<\xi<\half$.  
In appendix \ref{App:El} we report the elementary distributions for
the process \eqref{eq:El-proc} as functions $\xi$ and $x$. They depend
on the ``configurations'' of the three partons. If we consider only
photon exchange it is enough to identify the gluon momentum, we use
the index $\conf=1,2$ or $3$ to denote that the gluon momentum is
$P_1,P_2$ or $P_3$ respectively.  If we consider also $Z_0$ exchange,
with parity-violating terms, we need to identify also the nature of
incoming parton, this will be identified by the index $\can=q,\bar
q,g$. Finally for quarks we need to introduce also the flavour index
$f$. The configuration of the hard vertex will be then defined by the
configuration index $\rho=\{\conf,\can,f\}$, see appendix \ref{App:El}
for a detailed presentation of these conventions.

\subsection{The observable at parton level}
The QCD process involving multi-parton emission is described by one
incoming parton of momentum $p_1$ (inside the proton) and two outgoing
hard partons $p_2,p_3$ accompanied by an ensemble of secondary partons
$k_i$
\begin{equation}
  \label{eq:partonproc}
  q\,p_1 \to p_2\,p_3\,k_1\cdots k_n\>,\qquad p_1=x_1 P\,.
\end{equation}
Here, taking a small subtraction scale $\mu$ (smaller than any other
scale in the problem), we have assumed that $p_1$ (and the spectators)
are parallel to the incoming proton. 

Since $H(\chi)$ in \eqref{eq:H} is linear in $p_h$ and $p_{h'}$ we may
replace the sum over hadrons with a sum over outgoing partons for the
process \eqref{eq:partonproc}.  
To compute the distribution for small $\chi$ at SL accuracy, we can
consider the secondary partons to be soft and take the contributions
to $H(\chi)$ up to the first linear correction in the soft limit. 
Hard collinear radiation in the jets gets embodied into the hard scale
of the corresponding radiator, see \cite{PTstandard}, and, in the case
of the incoming jet, contributes to the parton density function, see
\cite{disko,hh,TDIS}.
In this limit the outgoing hard partons $p_2$ and $p_3$ tend to $P_2$
and $P_3$, with soft recoil we need to consider to first order. From
appendix \ref{App:obs}, we have
\begin{equation}
  \label{eq:recoil}
\begin{split}
  p_{t2}p_{t3} \simeq P^2_t\left(1\!-\!\sum_i\frac{|k_{yi}|}{P_t}\right),
  \qquad \chi_{23}=\pi\!-\!|\phi_{23}|\simeq |\phi_x|\>,
  \qquad \phi_x=\sum_i \frac{k_{xi}}{P_t}\,,
\end{split}
\end{equation}
with $k_{xi}$ the out-of-event-plane momentum component. The event
plane is defined in appendix \ref{App:obs} and $P_t$ in
\eqref{eq:Qab}.  Neglecting contributions to $H(\chi)$ from two
secondary soft partons we have (see figure~\ref{fig:ac_kin})
\begin{equation}
  \label{eq:H-parton}
\begin{split}
H(\chi)\!\simeq\! \frac{2P_t^2}{Q^2}
\left\{\!
\left(\!1\!-\!\sum_i\frac{|k_{yi}|}{P_t}\right)\!\de(\chi\!-\!\chi_{23})
\!+\!\sum_i\frac{k_{ti}}{P_t}
[\,\de(\chi\!-\!\chi_{2i})\!+\!\de(\chi\!-\!\chi_{3i})\,]\!\right\}.
\end{split}
\end{equation} 

\EPSFIGURE[ht]{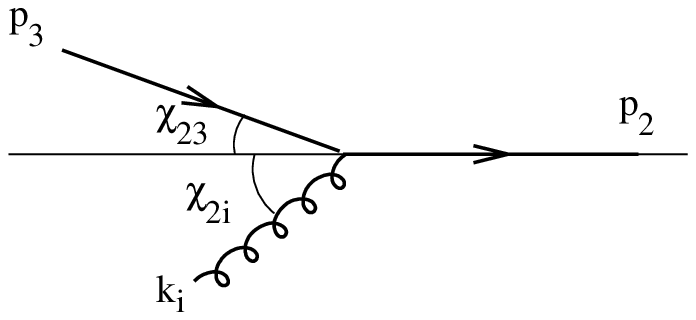,width=0.7\textwidth} {The kinematics of
azimuthal correlation in the plane orthogonal to the Breit axis.
The hard partons $p_2$ and $p_3$ correspond to the configuration
$\conf=1$ and $k_i$ is a soft secondary gluon.\label{fig:ac_kin}}

From the expression of $\chi_{ai}=\pi\!-\!|\phi_{ai}|$ we obtain in
the soft limit (see appendix \ref{App:obs})
\begin{equation}
  \label{eq:ia}
[\,\de(\chi\!-\!\chi_{2i})\!+\!\de(\chi\!-\!\chi_{3i})\,]\simeq
\de\left(\chi-|\bar\phi_i-\phi_x|\right),
\qquad |k_{yi}|\simeq k_{ti}\>|\cos\bar\phi_i|\>,
\end{equation}
with $\bar\phi_i$ the azimuthal angle between $k_i$ and either the
hard parton $p_2$ or $p_3$:
\begin{equation}
\label{eq:barphi}
\bar{\phi}_i=\left\{
\begin{split}
&\phi_{i2} \mbox{ for $k_i$ near $p_2$,} \\
&\phi_{3i} \mbox{ for $k_i$ near $p_3$.}
\end{split}
\right.
\end{equation}
Splitting the expression of $H(\chi)$ into a `hard' and a `soft'
contribution we can write, to first order in softness,
\begin{equation}
H(\chi) \simeq H_h(\chi)\>+\>H_s(\chi)\>,
\end{equation}
where, 
\begin{equation}
\label{eq:Hh,Hs}
\begin{split}
& H_h(\chi) = \frac{2P_t^2}{Q^2}\,\de(\chi-|\phi_x|)\>,\\
& H_s(\chi) = \frac{2P_t^2}{Q^2}\sum_i \frac{k_{ti}}{P_t}\left[
\delta\left(\chi-\left|\bar{\phi}_i-\phi_x\right|\right)
-|\cos\bar{\phi}_i|\,\delta(\chi-|\phi_x|)\right].
\end{split}
\end{equation}
This form explicitly shows that $H(\chi)$ is, by itself, a CIS
observable in the sense that the azimuthal correlation has nothing
more than the singularities of the total DIS cross section, namely the
incoming parton collinear singularities leading to the anomalous
dimension, see later.
Indeed, the hard term $H_h$ (hard-hard contribution with no recoil,
$p_{t2}p_{t3}\to P_t^2$) depends only on the total recoil $|\phi_x|$
from secondary partons.
The soft term $H_s$ (hard-soft contribution together with the recoil
piece of hard-hard contribution) is present only in the real
contribution.  Here the {\em soft}\ singularity of the matrix element 
is damped by the $k_{ti}$ factor, while the {\em
  collinear}\ singularity, $\bar\phi_i\!\to\!0$, is regularised by the
vanishing difference of the delta functions in the square brackets.

\section{Resummation}
Considering the region $\chi\ll 1$, the starting point for the QCD
resummation of the azimuthal distribution is the factorization
\cite{DDT-BCM} of the square amplitude for the process
\eqref{eq:partonproc} with $n$ secondary partons soft or collinear to
the primary partons $P_a$ in the hard vertex \eqref{eq:El-proc}. For each
configuration $\rho$ we have
\begin{equation}
  \label{eq:Mn}
  |M_{n}(k_1\ldots k_n)|^2\simeq|M_{0}|^2\cdot S_{n}(k_1\ldots k_n)\>.
\end{equation}
The first factor is the squared amplitude for the elementary vertex
\eqref{eq:El-proc} for the different configurations $\rho$. It gives
rise to the elementary hard distribution $d\hat \sigma_{\rho}$ as
function of the kinematical variables $x$ and $\xi$ in
\eqref{eq:El-kin}, see appendix \ref{App:El}. The second factor is the
distribution of secondary partons emitted from the hard momenta $P_a$.
Since in the emission, the nature of incoming parton changes, $S_{n}$
is a matrix in the configuration index $\rho$.  Corrections to the
factorized expression are relevant for finite $\chi$ and can be taken
into account by a non-logarithmic coefficient function.  From
\eqref{eq:Mn} the azimuthal correlation \eqref{eq:cH} is then given by
\begin{equation}
\label{eq:QCD-dsigma}
\begin{split}
\frac{d\sigma(\chi,y_{\pm})}{d\chi\,d\xB\, dQ^2}\>=\>
\sum_{\rho}\int_{\xB}^{x_M}\frac{dx}{x}
\int_{\xi_-}^{\xi_+}d\xi
\left(\frac{d\hat\sigma_{\rho}}{dx\,d\xi\,dQ^2}\right)\>C_{\rho}(\as)
\cdot \cI_{\rho}(\chi;Q_{ab})\>,
\end{split}
\end{equation}
where the limits $x_M$, $\xi_{\pm}$ are determined by the values of
$y_{\pm}$ which limit the transverse momenta, see appendix
\ref{App:El}.  Here $C_{\rho}(\as)$ is the non-logarithmic coefficient
function
\begin{equation}
  \label{eq:C}
  C_{\rho}(\as)=1+\frac{\as}{\pi}\,c_{1}^{(\rho)}(\chi)+\cO{\as^2}\>.
\end{equation}
The distribution $\cI$ depends on the geometry of the dijet system
through the hard scales $Q_{ab}$ given in \eqref{eq:Qab}. It is given
by
\begin{equation}
  \label{eq:cI}
\cI(\chi)=\!\int_0^1\!dx_1\cP\left(x_1,\mu\right)\>
\sum_n\frac{1}{n!}\int d\Gamma_n\cdot S_n(k_1\ldots k_n)\>,
\end{equation}
where $\cP(x_1,\mu)$ are the parton density functions at the (small)
subtraction scale $\mu$ and $x_1$ is the incoming parton momentum
fraction \eqref{eq:partonproc}. The phase space and observable are 
contained in
\begin{equation}
\label{eq:dGamma}
d\Gamma_n=\prod_{i=1}^n\int\frac{d^3k_i}{\pi\om_i}\>
\cdot \left\{H_h(\chi)+H_s(\chi)\right\}\cdot
\de\left(\frac{\xB}{x}\!-\!x_1\prod_{i\in\cC_1}z_i\right)\,.
\end{equation}
The last delta function accounts for the fact that the incoming parton
$p_1$ loses energy by radiation. Here $z_i$ is the collinear splitting
energy fractions for radiation of secondary partons $k_i$ in the
region $\cC_1$ collinear to $p_1$.  Note that it is this last delta
function which is responsible for the non cancellation of the
(residual) incoming parton collinear singularities which build up the
parton density function.

The resummation of the distribution $\cI$ is based on the
factorization of soft and collinear contributions in $S_n$. Then one
needs to factorize also the expression of $d\Gamma_n$.  For the
observable $H(\chi)$ this is done by $1$-dimensional Fourier transform
and, for the delta function in $\xB$, by Mellin transform
\begin{equation}
\label{eq:HFourier}
\begin{split}
&H_h(\chi)\!=\!\frac{2P_t^2}{Q^2}
\!\int_{-\infty}^\infty\frac{P_tdb}{\pi}
\cos(b\chi P_t)\prod_i e^{ib k_{xi}}\>,\\
&H_s(\chi)= \frac{2P_t^2}{Q^2}
\!\int_{-\infty}^\infty\frac{P_tdb}{\pi}
\cos(b\chi P_t)\prod_i e^{ib k_{xi}}
\left\{\sum_j\frac{k_{tj}}{P_t}
\left(e^{-iP_tb\bar\phi_j}\!-\!|\cos\bar\phi_j|\right)\right\},\\
&\de\left(\frac{\xB}{x}\!-\!x_1\prod_{i\in\cC_1}z_i\right) = 
\frac{1}{x_1}\int\frac{dN}{2\pi i}\left(\frac{\xB}{xx_1}\right)^{-N}
\prod_i\eps(z_i)\>,
\end{split}
\end{equation}
where the last integral runs parallel to the imaginary axis with
Re$(N)\!>\!1$ and
\begin{equation}
  \label{eq:eps}
\eps(z_i)=z_i^{N-1}\!,\>\>\mbox{for}\>\> k_i\in \cC_1\,,\qquad
\eps(z_i)=1,\>\> \mbox{for}\>\> k_i\in\!\!\!\!\!/\> \cC_1\,.
\end{equation}
Each secondary parton contributes to $d\Gamma_n$ by a factor
\begin{equation}
  \label{eq:source}
[\eps(z_i)\cos (bk_{xi})\!-\!1]\>=\>
[\eps(z_i)\!-\!1]\cos (bk_{xi})\>-\>[1\!-\!\cos (bk_{xi})]\>,
\end{equation}
with the one included to take into account virtual corrections. Since
$S_n$ is even in all $k_{xi}$, we have replaced $e^{ibk_{xi}}
\to\cos(bk_{xi})$.

The splitting made in \eqref{eq:source} shows that we can resum
independently the parton density function and the radiation function of
the observable $H$. Indeed, from the well known SL approximation
\begin{equation}
  \label{eq:theta}
  [1\!-\!\cos(bk_{xi})]\>\simeq\> 
\Theta\left(|k_{xi}|\!-\!\bar b^{-1}\right)\>,\qquad
\bar b=e^{\gam_E}\,|b|\>,
\end{equation}
we have that the first term in \eqref{eq:source} involves partons in
the region $k_{xi}\la \bar b^{-1}$ collinear to $P_1$ with weight
$[z_i^{N\!-\!1}\!-\!1]$. Their contributions reproduce the anomalous
dimension and resum to the parton density function at the proper hard
scale which is $\bar b^{-1}$.
The second term in \eqref{eq:source} involves partons in the
complementary region with weight $-1$, only virtual contributions
survive. Their contributions are resummed to give the standard
radiation factor for the observable $H(\chi)$ with (virtual)
frequencies larger than $\bar b^{-1}$.  This method to show the
factorization of these two contributions has been used and discussed
in \cite{disko,hh,TDIS}.

The final result for $\cI$ thus becomes
\begin{equation}
\label{eq:cI-fine}
\cI_{\can,\conf,f}(\chi;Q_{ab})
=\frac{4P_t^2}{Q^2}\int_{0}^\infty
\frac{P_tdb}{\pi}\cos(b\chi P_t)
\cdot\cP_{\can,f}\left(\frac{\xB}{x},\bar b^{-1}\right)\cdot
\cA_{\conf}(b;Q_{ab})\>,
\end{equation}
where the radiation factor is given by the sum of the hard and the
soft contribution
\begin{equation}
  \label{eq:cA}
\cA_{\conf}(b)=\cA_{\conf}^{h}(b)+\cA_{\conf}^{s}(b)\>,\qquad
\cA_{\conf}^{h}(b)=e^{-\cR_{\conf}(b)}\>,\quad
\cA_{\conf}^{s}(b)=e^{-\cR_{\conf}(b)}\cdot B_{\conf}(b)\>.
\end{equation}
It will be seen that only the hard contribution is required for a PT
resummation at SL level, while the soft contribution is responsible
for leading NP corrections.

Here $\cR_{\conf}$ is the full CIS radiator for the secondary emission
and is given by
\begin{equation}
  \label{eq:cR}
\cR_{\conf}(b) = \int \frac{d^3k}{\pi\om}\,
W_{\conf}(k)\,\left[1-\cos(bk_x)\right]\,.
\end{equation}
The quantity $B_{\conf}$, representing the $H_s(\chi)$ contribution to the
observable, is
\begin{equation}
\label{eq:B}
B_{\conf}(b)=\int\frac{d^3 k}{\pi\omega}W_\conf(k)\,\frac{k_t}{P_t}
\left[\cos(bP_t\bar{\phi})-|\cos\bar{\phi}|\right]\>.
\end{equation}
Here $W_{\conf}(k)$ is the
two-loop distribution for the emission of the soft gluon from the
primary three hard partons $P_a$ in the configuration $\conf$:
\begin{equation}
\label{eq:W}
W_{\conf}(k) = \frac{N_c}{2}\left(w_{\conf a}+w_{\conf b}-
\frac{1}{N_c^2}w_{ab}\right)\>,\qquad a\neq b\neq \conf\>,
\end{equation}
where $w_{ab}(k)$ is the standard distribution for emission of a soft
gluon from the $ab$-dipole.  It has a simple expression in the
$ab$-dipole centre of mass.  Representing in this frame the soft gluon
momentum by Sudakov decomposition
\begin{equation}
  \label{eq:absud}
  k^*=\al P_a^*+\be P_b^*+\ka\>,\qquad \al\be=\frac{\ka^2}{Q_{ab}^2}\>, 
\end{equation}
with $P_a^*$ and $P_b^*$ back-to-back momenta ($Q^2_{ab}\!=\!2P_a^*P_b^*$)
and $\ka$ a two-dimensional transverse vector, we can write
\begin{equation}
  \label{eq:wab}
  w_{ab}(k)=\frac{\as(\ka)}{\pi \ka^2}\>,\qquad
\frac{d^3k}{\pi \om}=\frac{d^2\ka}{\pi}\frac{d\al}{\al}\>, 
\quad \al>\frac{\ka^2}{Q_{ab}^2}\>.
\end{equation}
To ensure SL accuracy, we have to take $\as$ defined in the physical
scheme \cite{CMW}.  Taking $P_a^*,P_b^*$ along the $z$-axis, the
$x$-components of the momenta are the same as the ones in the
laboratory frame with the event plane given by the $yz$-plane. We then
have $k_x\!=\!\ka_x$.

A similar analysis gives the (weighted) total cross section
$\sigma(y_{\pm})$ in \eqref{eq:Sigma} needed to obtain the normalized
differential distribution $d\Sigma(\chi)/d\chi$. We find
\begin{equation}
\label{eq:sigma}
\sigma(y_{\pm})\>=\>\sum_{\rho}\int_{\xB}^{x_M}\frac{dx}{x}
\int_{\xi_-}^{\xi_+}d\xi
\left(\frac{d\hat\sigma_{\rho}}{dx\,d\xi\,dQ^2}\right)
\frac{4P_t^2}{Q^2}\cP_{\rho}\left(\frac{\xB}{x},Q\right)\>,
\end{equation}
where the parton density function is reconstructed at the
unconstrained DIS hard scale $Q$. 

In conclusion we have that, while in the DIS cross section the proper
hard scale of the parton density functions is $Q$, in the distribution
\eqref{eq:cI-fine} it is given by $\bar{b}^{-1}$. This is a result of
QCD coherence.

\section{The result}
The general structure in \eqref{eq:cA} of the radiation factor is
similar to the one in EEC: the hard term $\cA_{\conf}^{h}$ gives the
PT contribution at SL accuracy. The soft term $\cA_{\conf}^{s}$ is
beyond SL accuracy. On the contrary, for the NP correction, it is the
soft term which gives the leading $1/Q$-power piece; the hard term
gives a subleading $1/Q^2$ piece.  We discuss these results in the two
next subsections.

\subsection{Hard contribution}
We start from the PT part of the radiator \eqref{eq:cR} which is given
at SL accuracy by the approximation \eqref{eq:theta}. This has been
computed in \cite{disko} (where it corresponds to the replacements
$\nu\to0$ and $\nu\be_2,\nu\be_3\to b$).  We have
\begin{equation}
\begin{split}
\label{eq:PT-rad}
\cR^{\PT}_{\conf}(b) \!=\!R_{\conf}(b) 
\!=\! \sum_{a=1}^3 C_a^{(\conf)} 
r\left(\bar b, \zeta_a^{(\conf)}Q_a^{(\conf)}\right)\>, \qquad
r\left(\bar b,Q\right) \!=\! \int_{1/\bar b}^Q\frac{dk}{k}
\frac{2\as(2k)}{\pi}\ln\frac{Q}{2k}\>,
\end{split}
\end{equation}
where the various hard scales and constants are, see \cite{disko,hh},
\begin{equation}
  \label{eq:QCs}
\begin{split}
&Q_{a}^{(\conf)}=Q_{b}^{(\conf)}=Q_{ab}\>,\qquad
C_{a}^{(\conf)}=C_{b}^{(\conf)}=C_F\>,\qquad
\zeta_{a}^{(\conf)}=\zeta_{b}^{(\conf)}=e^{-\frac{3}{4}}\>,\\
&Q_{\conf}^{(\conf)}=\frac{Q_{a\conf}Q_{\conf b}}{Q_{ab}}\>,\qquad\quad\>
C_{\conf}^{(\conf)}=C_A\>,\qquad\qquad\quad\>
\zeta_{\conf}^{(\conf)}=e^{-\frac{\beta_0}{4N_c}}\>,
\end{split}
\end{equation}
with $\conf$ a gluon and $a$ or $b$ a quark (or antiquark), $\be_0$ is
the first coefficient of the beta function (see \eqref{eq:K-beta0}).
As is typical for a $3$-jet quantity, the scale for the quark or
antiquark terms of the radiator is the $\qq$ invariant mass, while the
scale for the gluon term is given by the gluon transverse momentum
with respect to the $\qq$ system. The rescaling factors
$\zeta_{a}^{(\conf)}$ take into account SL contributions from non-soft
secondary partons collinear to the primary partons $P_a$.
The integration variable $k$ in \eqref{eq:PT-rad} is the out-of-plane
component of momentum $k_x$ and $\as$ is taken in the physical scheme
\cite{CMW}. The rescaling factor $2$ in the running coupling comes
from the integration over the in-plane momentum component. The exact
expression of hard scales and of rescaling factors in \eqref{eq:QCs}
is relevant at SL level.

The complete SL resummed result is then given by
\begin{equation}
\label{eq:PT-fine1}
\begin{split}
\frac{d\Sigma^{\PT}(\chi,y_{\pm})}{d\chi}= \sigma^{-1}(y_{\pm})
\sum_{\rho}\int_{\xB}^{x_M}\frac{dx}{x}\int_{\xi_-}^{\xi_+}d\xi
\left(\frac{d\hat\sigma_{\rho}}{dx d\xi dQ^2}\right)\>
C_{\conf}(\as)
\cdot \cI^{\PT}_{\rho}(\chi) \>,
\end{split}
\end{equation}
where 
\begin{equation}
\label{eq:PT-fine2}
\begin{split}
\cI^{\PT}_{\rho}(\chi)\!=\!
\int_{0}^\infty \frac{P_tdb}{\pi}\cos(b\chi P_t)\>\cF_{\rho}(b)\,,\quad 
\cF_{\rho}(b)\equiv \frac{4P_t^2}{Q^2}\>
\cP_{\can,f}\left(\frac{\xB}{x},\bar b^{-1}\right)\>
e^{-R_{\conf}(b)}\,.
\end{split}
\end{equation}
The integrand is computed at SL accuracy in $b$-space, that is,
$\ln\cF_{\rho}(b)$ resums all terms $\as^n\ln^{n+1}bQ$ and
$\as^n\ln^{n}bQ$.  For $\chi\to0$ we have that $\cI^{\PT}(0)$ has a
$1/\sqrt{\as}$-singular behaviour (it is essentially given by the
integral of a Sudakov form factor, see \cite{PP} for the EEC case).
Formally the upper limit of $b$-integration is set to infinity, but
actually, from \eqref{eq:PT-rad} is $\bar b<2/\LQCD$.

As we shall see, the differential azimuthal distribution is
monotonically decreasing by increasing $\chi$. This is in contrast
with the EEC case in which the differential distribution (analogous to
$d\Sigma(\chi)/d\chi$) develops a peak in the small $\chi$ region (the
back-to-back polar angle in this case), which is present also in the
data (see \cite{EECdata}).  In the next section we present the
numerical evaluation of the azimuthal distribution while for the
remainder of this section we discuss the origin of this behaviour and
of the differences with respect to the EEC case.

\paragraph{General features}
To discuss $\cI^{\PT}(\chi)$ for small $\chi$ (index $\conf$ neglected
here) we divide the $b$-integral at the point $b_0=1/(\bar{\chi}P_t)$
with $\bar{\chi}=\chi e^{\gam_E}$:
\begin{equation}
  \label{eq:cIpm}
\cI^{(-)}(\chi)=\frac{d}{d\chi}\int_0^{b_0}\frac{db}{\pi b}
\sin(\chi b P_t)\> \cF(b)\>,\qquad
\cI^{(+)}(\chi)=\frac{d}{d\chi}\int_{b_0}^{\infty}\frac{db}{\pi b}
\sin(\chi b P_t)\>\cF(b)\>.
\end{equation}
The sine function oscillates with a first maximum above the splitting
point.  The integrand $\cF(b)$ has a Sudakov behaviour at large $b$
(see in appendix~\ref{App:sud} the discussion for $\cF$ in DL
approximation).  From this behaviour we have that the two
contributions behave as follows:
\begin{itemize}
\item $\cI^{(+)}(\chi)$ is dominated by the lower bound at $b=b_0$
\begin{equation}
\label{eq:cIsud}
\cI^{(+)}(\chi)\>\simeq\> 
\frac{e^{-R\left(1/\bar{\chi}P_t\right)}}{\chi}\cdot\cS^{(+)}(\chi)\>,
\end{equation}
where, for small $\chi$, the remainder $\cS^{(+)}(\chi)$ can be
expanded in the SL function
\begin{equation}
  \label{eq:R'}
  R'(\chi)=2C_T\frac{\as}{\pi}\ln\frac{1}{\chi}\>,
\quad C_T=2C_F+C_A\>.
\end{equation}
The Sudakov factor $e^{-R}/{\chi}$ in \eqref{eq:cIsud} has a peak at
$\chi\sim\chi_1$ with $R'(\chi_1)=1$ which corresponds to
\eqref{eq:peak}.  This Sudakov behaviour is expected since large
values of $b$ force all the emitted $k_{xi} $ to be soft, $|k_{xi}|\la
\chi P_t$.
\item $\cI^{(-)}(\chi)$ takes contributions from values of $b$ not
  necessarily large. It is a decreasing function of $\chi$ from the
  value $\cI^{(-)}(0)$ and has a $1/\sqrt{\as}$-singular behaviour.
  For $b$ not large, the soft recoil $\chi P_t=|\sum_i k_{xi}|$ is
  obtained by cancellations among some larger $k_{xi}$.
\end{itemize}
The relative size of the two contributions in \eqref{eq:cIpm} can be
estimated by the DL approximation of $\cF$, see
appendix~\ref{App:sud}. One finds that the two terms
$\cI^{(\pm)}(\chi)$ become comparable at $\chi\sim \chi_1$, i.e. just
near the Sudakov peak and for smaller $\chi$ the contribution
$\cI^{(-)}(\chi)$ takes over.  As a result, the differential azimuthal
distribution $d\Sigma(\chi)/d\chi$ continuously decreases from
$\chi=0$ and does not have a peak at small $\chi$.

\paragraph{Comparison with EEC.}
The behaviour of $d\Sigma_{\ee}(\chi)/d\chi$, the differential EEC
with $\chi$ the back-to-back polar angle in $\ee$, is very different.
This distribution is given by an expression similar to
\eqref{eq:PT-fine2} with the most important difference\footnote{In EEC
  case the integrand $\cF(b)$ is simply given by the exponential of a
  radiator $R(b)$, which involves only two hard emitters, the $\qq$
  pair with total charge $C_T=2C_F$, and the cosine is replaced by the
  Bessel function.}  being that in EEC the integration measure is
$\chi d^2b$ instead of $db$.  As a consequence, the contribution
involving the finite $b$ regions, 
which corresponds to $\cI^{(-)}(\chi)$ in \eqref{eq:cIpm}, 
is relevant\footnote{From the DL approximation this term becomes
  comparable to the large $b$-term, corresponding to $\cI^{(+)}(\chi)$
  in \eqref{eq:cIpm}, at $\chi\sim \chi_2=\chi^2_1$, i.e.
  $R'(\chi_2)=2$. See \cite{PP}.}  only for
$\chi\!\ll\!\chi_1$ with $\chi_1$ the position of the Sudakov peak
given in \eqref{eq:peak} for $C_T=2C_F$.  Therefore, the distribution
$d\Sigma_{\ee}(\chi)/d\chi$ manifests a Sudakov behaviour in the small
$\chi$ region.  The fact that the contribution involving the finite
$b$-region is essentially negligible in the EEC case can be argued by
observing that it is dominated by emitted transverse momenta
$\vec{k}_{ti}$ not necessarily soft which have to sum up to a total
soft recoil. The constraint of momentum cancellation in $2$-dimension
is much stronger than in $1$-dimension as for the azimuthal case.

A similar way to see the different behaviour in the EEC and \EET\ 
cases, is obtained by trying to factorize the Sudakov form factor
$e^{-R}/\chi$ from the differential distribution. In the EEC case the
remainder factor can be expressed in terms of the SL variable
$R'(\chi)$ (with $C_T=2C_F$) in the full range $\chi\ga\chi^2_1$ (with
$\chi_1$ the position of the Sudakov peak).  In the \EET\ case instead
the remainder factor is a function of $R'(\chi)$ only in the range
$\chi\ga\chi_1$ while at smaller $\chi$ it does cancel the
$e^{-R}/\chi$ factor and then it is not any more a SL function, see in
appendix \ref{App:sud} the analysis in the DL approximation.

\subsection{Full contribution}
To complete the analysis we need to take into account also
$\cA_{\conf}^{s}$ in \eqref{eq:cA} corresponding to $H_s(\chi)$, the
soft contribution to the observable, see \eqref{eq:Hh,Hs}. From
\eqref{eq:W} we can write
\begin{equation}
  \label{eq:B'}
B_{\conf}(b) = \frac{N_c}{2}\left(B_{\conf a}+B_{\conf b}-
\frac{1}{N_c^2}B_{ab}\right)\>,\qquad a\neq b\neq \conf\>,  
\end{equation}
with the $\{ab\}$-dipole contribution given by
\begin{equation}
\label{eq:Bab}
B_{ab}(b)=\int_0^{Q_{ab}}\frac{d^2 \ka}{\pi\ka^2}\frac{\as(\ka)}{\pi}\,
\int^{1}_{\ka^2/Q^2_{ab}}\frac{d\al}{\al}\cdot \frac{k_t}{P_t}
\left[\,\cos(bP_t\bar{\phi})-|\cos\bar{\phi}\,|\right]\>.
\end{equation}
Here the integration variables $\ka$ and $\al$ are the momentum
components of the soft gluon in the $\{ab\}$-dipole centre of mass
(see \eqref{eq:absud} and \eqref{eq:wab}), while $k_t$ and $\bar\phi$
are the transverse momentum and azimuthal angle in the Breit frame,
see \eqref{eq:barphi}.  In appendix \ref{App:Bsoft} we give their
kinematical relations. For instance, for the 12-dipole we have
\begin{equation}
  \label{eq:12}
  k_y=\al P_t+\ka_y\>,\qquad \tan\bar\phi=\frac{\ka_x}{k_y}\>,
\end{equation}
where we used the fact that the Lorentz transformations between the
two frames do not involve the $x$-direction so that $k_x=\ka_x$.  The
result \eqref{eq:12} can be easily recovered for instance observing
that for soft transverse momentum in the dipole frame $\ka\to0$, the
momentum $k$ becomes collinear to $P_2$ for $\al>0$ and to $P_1$ for
$\al\to0$.

The two terms in the square brackets of \eqref{eq:Bab} have DL
singularities for $\ka\to0$ and $\bar\phi\to0$. However the sum (multiplied
by $k_t$) is regular which gives
\begin{equation}
  \label{eq:cAs}
  \cA_{\conf}^{s}(b)\sim \as \cdot \cA_{\conf}^{h}(b)\>,
\end{equation}
so that this term contributes at PT level beyond SL accuracy, see
appendix \ref{App:Bsoft} and \cite{EEC}. This fact is carefully
explained at the end of section~\ref{sec:QCD}. Recall that the cancellation
is between the contributions to the observable $H(\chi)$ from the recoil
of the two outgoing hard partons $\{p_2p_3\}$ and from a soft secondary 
and a hard parton $\{k_ip_{2/3}\}$.

The soft part of the radiation factor $\cA_{\conf}^{s}$ gives instead
the leading NP correction which is proportional to $b$, see also
\cite{EEC}.  This NP correction comes from the region in which the
soft gluon in $B_{ab}$ is collinear to one of the outgoing hard partons
$P_2$ or $P_3$. For instance for the $B_{12}$ radiator, where
\eqref{eq:12} holds, this region corresponds to
\begin{equation}
  \label{eq:coll2}
  k_y\simeq\al P_t\gg \ka_x\>,
\end{equation}
which gives
$$
k_t\simeq \al P_t\>,\qquad
\cos (bP_t\,\bar\phi)\simeq\cos\,\frac{b\ka_x}{\al}\>,\qquad
\cos\bar\phi\simeq1\>.
$$
Changing variable $u=b|\ka_x|/\al$, and taking in the integrand the
leading piece for $\ka\to0$, the contribution to $B_{12}(b)$ from this
region is
\begin{equation}
  \label{eq:deB21}
\begin{split}  
\de B_{12}(b)&\!=\!b\!\int\frac{d\ka^2}{\ka^2}\frac{\ka\,\as(\ka)}{\pi}
\!\int_{-\pi}^{\pi}\frac{d\phi}{2\pi}|\cos\phi|
\!\int_0^\infty\!\frac{du}{u^2}[\cos u\!-\!1]
=-b\cdot\!\int\frac{d\ka^2}{\ka^2}\frac{\ka\,\as(\ka)}{\pi}\,,
\end{split}
\end{equation}
with the integration over the NP region of small $\ka$.
It is simple to see that the regions non-collinear to the outgoing
hard parton $P_2$ (away from \eqref{eq:coll2}) lead to contributions
which are even in $b$ and then give subleading NP corrections. A
correction linear in $b$ is non-local in the conjugate variable
$\chi$, see \cite{EEC}.

To complete the calculation of the leading NP correction we use the
standard procedure extensively discussed in \cite{EEC,milan}.  First
we need to extend the meaning of the running coupling into the large
distance region. This can be done by following the approach \cite{DMW}
in which one uses the dispersive representation
\begin{equation}
  \label{eq:dmw}
  \frac{\as(\ka)}{\ka^2}=
\int_0^{\infty}\!dm^2\,\frac{\aef(m^2)}{(\ka^2+m^2)^2}\,.
\end{equation}
Moreover one needs to take into account the non fully inclusive
character of the observable $H(\chi)$. This can be done in two steps,
see \cite{milan}. One replaces the momentum $\ka$ of the emitted
soft gluon by $\ka\to \sqrt{\ka^2+m^2}$, with $m$ the mass of the
gluon decaying products. Then one multiplies the result by the Milan
factor $\cM$.  Finally one obtains
\begin{equation}
  \label{eq:deB21'}
\de B_{12}(b)=-b\cdot\cp\>,\qquad 
\cp=\frac{2\cM}{\pi}\int dm\,\daef(m^2)\>,
\end{equation}
with $\daef$ the large distance contribution of the effective
coupling.  See appendix \ref{App:Bsoft} for the connection of $\cp$
with the NP parameter already measured in $2$-jet observables.

Similar leading NP corrections are obtained for the $B_{13}$ and
$B_{23}$ pieces. Since they come from the phase space regions
collinear to the outgoing hard partons $P_2$ and $P_3$, the complete
NP correction from \eqref{eq:B'} is proportional to the total charge
of the hard outgoing primary partons
\begin{equation}
  \label{eq:dB}
\de B_{\conf}(b) = 
-\left(C_2^{(\conf)}\!+\!C_3^{(\conf)}\right)\,b\cdot\cp\>.
\end{equation}

In the present analysis we concentrate on the leading power correction
discussed above. Large angle emissions and the ones collinear to the
incoming parton generate next-to-leading NP corrections, proportional
to $b^2\sim Q^{-2}$.  They appear not only in this $B$ term, but also
in the radiator $\cR$ (see \eqref{eq:cR}) and in the parton density
functions $\cP$ (see \cite{DMW}).  A complete study of higher orders
NP corrections is still to be done.

The final answer for $\cI$, including the leading NP correction in
\eqref{eq:dB}, thus is
\begin{equation}
\label{eq:fine1}
\begin{split}
&\cI_{\rho}(\chi;Q_{ab})=
\int_{0}^\infty\frac{P_tdb}{\pi}\cos(b\chi P_t)\>\cF_{\rho}(b)\cdot
\left\{1-\left(C_2^{(\conf)}\!+\!C_3^{(\conf)}\right)\,b\cdot\cp\right\},
\end{split}
\end{equation}
with $\cF_{\rho}(b)$ the PT factor in \eqref{eq:PT-fine2}.  The final
answer for the azimuthal correlation in \eqref{eq:Sigma} can be
written in the form
\begin{equation}
  \label{eq:fine2}
  \frac{d\Sigma(\chi,y_{\pm})}{d\chi}=
  \frac{d\Sigma^{\PT}(\chi,y_{\pm})}{d\chi}
\left\{1-\cp\VEV{(C_2+C_3)\cdot b}\right\},
\end{equation}
with $b$ (and the outgoing hard parton charges) averaged over the PT
distribution in \eqref{eq:PT-fine2}. One can give an estimation of
$\VEV{b}$ by using the one-loop coupling approximation. From appendix
\ref{App:gam} we evaluate this mean at $\chi=0$ and obtain
\begin{equation}
  \label{eq:gam}
\VEV{b}_{\chi=0}\sim \frac{1}{\LQCD}
\left(\frac{\LQCD}{Q}\right)^{\gam}\>, \qquad
\gam={\frac{4}{\be_0}C_T
\ln\frac{2+\frac{4}{\be_0}C_T}{1+\frac{4}{\be_0}C_T}}=0.62\>\>
(n_f\!=\!3)\>.
\end{equation}
Taking into account the two-loop coupling the exponent $\gam$ is not a
constant but smoothly dependent on $Q$, see \cite{EEC}. The detailed
numerical analysis will be done in the next section.

\section{Numerical evaluation \label{sec:num}}
In the small $\chi$ region, the PT distribution is obtained from
\eqref{eq:PT-fine1},\eqref{eq:PT-fine2} with the PT part of the
radiator in \eqref{eq:PT-rad}. At larger values of $\chi$ one should
use the exact PT result.  In order to obtain a PT description valid
for small and large $\chi$ one needs to match the resummed and the
fixed order distribution. This amounts in computing, to the due order,
the coefficient function $C_{\conf}(\as)$ in \eqref{eq:C}. This is
usually done by comparing the resummed expression with the exact
result.

We start from the integrated resummed distribution
$\Sigma^{\PT}(\chi)$ given by
\begin{equation}
\label{eq:Sigma-ev}
\Sigma^{\PT}(\chi)=\sigma^{-1}(y_{\pm})
\sum_{\rho}\int_{\xB}^{x_M}\frac{dx}{x}\int_{\xi_-}^{\xi_+}\!\!d\xi
\left(\frac{d\hat\sigma_{\rho}}{dx d\xi dQ^2}\right)
\int_0^{\infty}\!\frac{db}{\pi b}\sin(bP_t\chi)\cF_{\rho}(b)\>,
\end{equation}
with $\cF_{\rho}(b)$ the PT factor in \eqref{eq:PT-fine2} and
$\sigma(y_{\pm})$ the Born cross section (see \eqref{eq:sigma}).
In the following we describe in detail the adopted procedure for the
first order matching using a fixed order numerical program such as
\cite{DISENT,DISfix}.  Notice that, in order to control the $\as$
scale, it is not sufficient to use first order matching but one needs
to take into account also the two-loop analysis.  To first order in
$\as$ the resummed distribution $\Sigma^{\PT}(\chi)$ in
\eqref{eq:Sigma-ev} reads
\begin{equation}
\label{eq:Sigma-ev-first}
\Sigma^{\PT}(\chi)= \frac{1}{2} \left(1+\frac{\as}{2\pi}\left(G_{12}
L^2+G_{11}(y_{\pm})L+c_1^{\res}(y_{\pm})\right)+\cO{\as^2}
\right),
\qquad L=-\ln\chi\>,
\end{equation}
with $G_{12}=-2C_T$, $G_{11}$ and $c_1^{\res}$ given in
\eqref{eq:g11-c1res} and $\as=\al_{\MSbar}(Q)$.  Here the factor $1/2$ comes from the fact
that the distribution \eqref{eq:Sigma-ev} does not include the
singular contribution proportional to $\delta(\chi-\pi)$.
Only with this singular contribution the distribution is normalized to
one.

Following \cite{EEC}, we match the resummed expression in
\eqref{eq:Sigma-ev} with the exact first order result
$\Sigma_1(\chi)$, which is here obtained using the numerical program
DISENT of ref.~\cite{DISENT}. We obtain
\begin{equation}
\label{eq:Sigma-mat}
\Sigma^{\PT}_{\mat}(\chi)=\frac12+\left(1+\frac{\as}{2\pi}(c_1-c_1^{\res})\right)
\left(\Sigma^{\PT}(\chi)
+\delta\Sigma(\chi)\right)\>,
\end{equation}
with $\Sigma^{\PT}(\chi)$ the resummed distribution in
\eqref{eq:Sigma-ev}, the coefficient
\begin{equation}
\label{eq:c1}
c_1=\lim_{\chi\to 0} \left(\Sigma_1(\chi)-G_{12}L^2-G_{11}L\right).
\end{equation}
and the matching correction
\begin{equation}
\label{eq:delta-Sigma}
\delta\Sigma(\chi)=\frac12\left(\frac{\as}{2\pi}(\Sigma_1(\chi)-
G_{12}L^2-G_{11}L-c_1)\right)\>.
\end{equation}
Including the NP corrections we have:
\begin{equation}
\label{eq:Sigma-full}
\Sigma(\chi)=\frac12 +\left(1+\frac{\as}{2\pi}(c_1- c_1^{\res})\right)
\left(\left(1-\lam^{\NP}\VEV{(C_2+C_3)\cdot b}\right)
\Sigma^{\PT}(\chi)
+\delta\Sigma(\chi)\right)\>,
\end{equation}
where $b$ is now averaged over the PT integrated distribution in
\eqref{eq:Sigma-ev}.  

Here we report the result for the azimuthal
correlation distribution with the following choices.  We used the
parton density function set MRST2001\_1 \cite{MRST} corresponding to
$\as(M_Z)=0.119$. We also set $\al_0(2\GeV)=0.52$ (see \eqref{eq:cp}),
a value obtained from the analysis of two-jet event shapes \cite{SZ}.

In figure~\ref{fig:x0.10q30y1.0} we plot the full azimuthal distribution
$d\Sigma/d\chi$ obtained from \eqref{eq:Sigma-full} for $\xB=0.1$,
$Q^2=900\GeV^2$, $y_-=1$ and $y_+=2.5$.  The center of mass squared
energy has been fixed at $s=98400\GeV^2$.  In order to simplify the
analysis (data are not yet available) we considered only the case of
exchanged photon (see for instance the
discussion in \cite{disko}).  

Together with this curve we plot the
fixed order PT distribution obtained from DISENT and the PT resummed
distributions obtained from \eqref{eq:Sigma-ev} and
\eqref{eq:Sigma-mat}.  We notice that the PT resummed curve follows
the fixed order one for not too small values of $\chi$. Then, while
the first order starts developing a singularity for $\chi\to 0$, the
resummed curve rises to a constant value. Actually the height of the
`plateau' is dramatically reduced after performing the first order
matching. This is due to the large coefficient $c_1$ appearing in
\eqref{eq:Sigma-mat} ($c_1\simeq -30$ in the present case).

The effect of the NP correction \eqref{eq:Sigma-full} is to further
deplete the PT curve, in analogy with the EEC case \cite{EEC}.
 
\EPSFIGURE[ht]{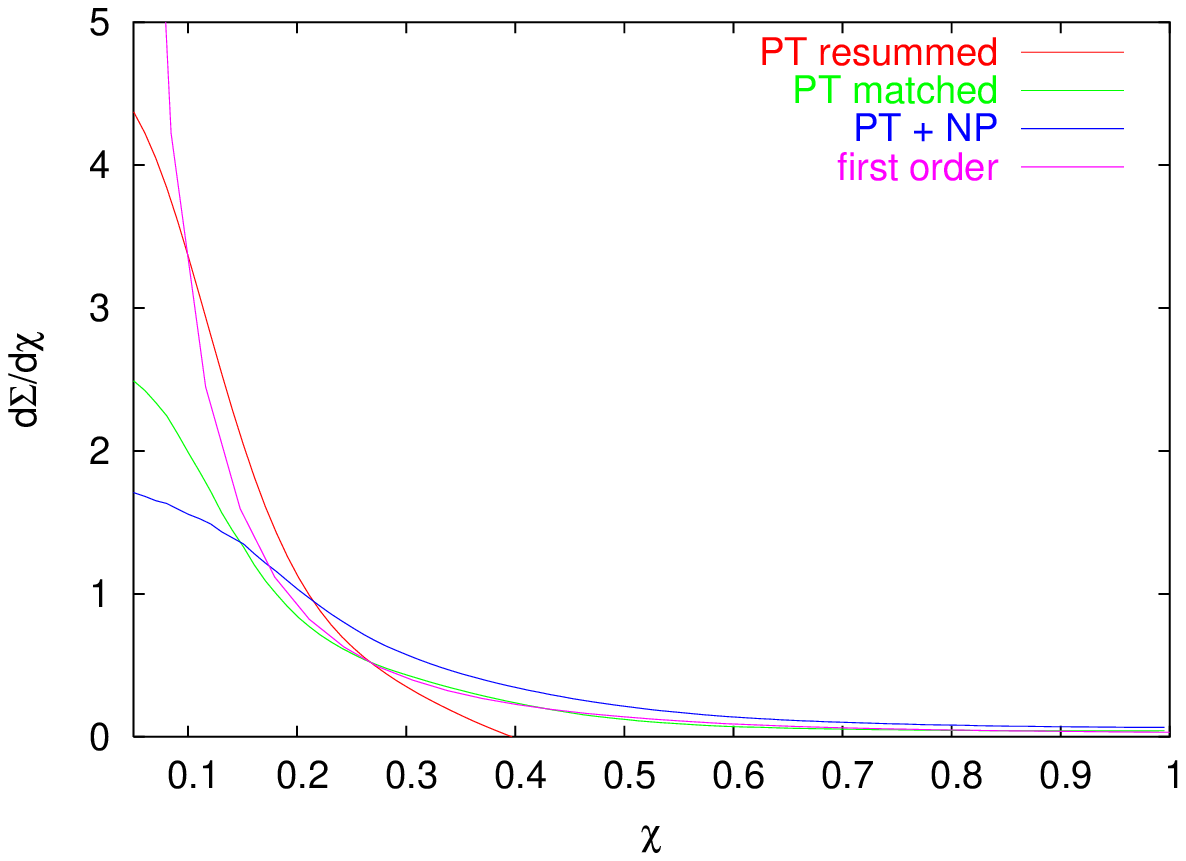,width=0.7\textwidth} {The azimuthal
  correlation distribution for $s=98400\GeV^2$, $\xB=0.1$, $Q^2=900\GeV^2$,
  $y_-=1.0$ and $y_+=2.5$. The upper curve is the resummed PT
  distribution obtained from (5.1) (without matching), the lower
  curves are the PT and NP matched predictions obtained from (5.3) and
  (5.6) respectively. In the figure is also shown the first order
  prediction given by DISENT.\label{fig:x0.10q30y1.0}}

\newpage
\section{Discussion}
We presented the QCD analysis of \EET, the azimuthal correlation in
DIS defined in \eqref{eq:H}-\eqref{eq:Sigma}.  We considered small
$\chi$, the back-to-back azimuthal angle, and finite $y_2$, the dijet
resolution variable which selects dijet events and limits emitted jet
transverse momenta not to exceed $Q$ so that the exchanged photon is
pointlike.  This distribution is the generalization to DIS of EEC in
$\ee$.

The QCD resummation is performed in the $1$-dimensional impact
parameter $b$ conjugate to the soft recoil $\chi P_t=|\sum_ik_{xi}|$,
with $k_{xi}$ the out-of-plane momenta of the undetected secondary
partons, see \eqref{eq:recoil}. We sum all $\as^n\ln^{n+1}\!bQ$ and
$\as^n\ln^{n}\!bQ$ terms in the exponent. The main complication with
respect to the $\ee$ case is that \EET\ involves also initial state
radiation which contributes both to the incoming parton evolution and
to the observable $H(\chi)$. Due to QCD coherence the two
contributions factorize, see \eqref{eq:cI-fine}.  We have indeed that
$\bar b^{-1}\sim \chi P_t$ sets both the hard scale for the parton
density functions $\cP$ (instead of $Q$ as in the cross section, see
\eqref{eq:sigma}) and the lower limit to frequencies in ``radiation
factor'' $\cA$, a CIS quantity associated to the $\qq g$ hard vertex.
See \cite{disko,hh,TDIS} for incoming radiation factorization in jet-shape
observables in processes involving incoming hadrons.

For the PT result at SL accuracy one needs to consider only the hard
term $H_h(\chi)$ of the observable from the two outgoing partons in
the $\qq g$ hard system, see \eqref{eq:Hh,Hs}.  
The general behaviour of $d\Sigma^{\PT}(\chi)/d\chi$ (and its comparison
with EEC) can be simply explained by observing that there are
essentially two phase space configurations for the secondary partons:
\begin{itemize}
\item all $|k_{xi}|\la\chi P_t$. This gives a contribution to
  $d\Sigma^{\PT}(\chi)/d\chi$ which behaves as a Sudakov form factor
  \eqref{eq:cIsud} with a peak at small $\chi$, see \eqref{eq:peak};
\item the other configurations, with some $|k_{xi}|$ exceeding the
  soft recoil. This corresponds to impact parameter with all values.
  Their contributions to $d\Sigma^{\PT}(\chi)/d\chi$ lead to a
  $1/\sqrt{\as}$-singular term (the integral of a Sudakov form factor)
  which is a decreasing function of $\chi$ starting from a finite
  value at $\chi=0$, see appendix~\ref{App:sud}.
\end{itemize}
The contributions from these two momentum configurations are
comparable just at the Sudakov peak and for smaller $\chi$ the
$1/\sqrt{\as}$-singular term is dominating.  As a result,
$d\Sigma^{\PT}(\chi)/d\chi$ is continuously decreasing with $\chi$.
Therefore the exponentiation of the one loop result in
\eqref{eq:Sig-1}, with the characteristic Sudakov peak, is valid only
above the Sudakov peak region.  This is qualitatively different from
the EEC behaviour in which the Sudakov peak is indeed manifest.

We have assumed \cite{DMW,NPstandard} that the leading power corrections
are obtained when the argument of the QCD coupling runs into large
distance regions.  Such a hypothesis has been successfully tested
in $2$-jet shape observables \cite{NPstandard} and EEC \cite{EEC}.
The azimuthal correlation (as other $3$-jet observables
\cite{3ee,disko,hh}) provides a further way to explore this
assumption.  Indeed, the running coupling enters our result with the
three different arguments $\vka_{ab}$, the transverse momentum in the
$ab$-dipole centre of mass frame. For each $ab$-dipole we evaluated
the leading power corrections by taking $\ka_{ab}\sim\LQCD$, thus
obtaining the result \eqref{eq:dB}.  The three soft regions
$\ka_{ab}\sim\LQCD$ are very different. Consider for instance the
leading power correction from the $12$-dipole which is coming from the
momentum region (see \eqref{eq:coll2})
\begin{equation}
  \label{eq:kka}
k_x=\ka_x\,,\quad |k_y|\gg |\ka_y|\sim |\ka_x|\sim\LQCD\>,
\end{equation}
where $\vka=\vka_{12}$ while $\vec{k_t}$ is the transverse momentum in
the Breit frame. We have that NP region $\ka\sim \LQCD$ does not
correspond to $k_t\sim\LQCD$.  One finds analogous differences
comparing $\vka_{12}$ with $\vka_{13}$ or $\vka_{23}$.
This shows that analyzing \EET\ (as well as $3$-jet observables) we
can further check whether NP corrections are controlled by the soft
arguments in running coupling.

The strength of the leading power correction is controlled by the NP
parameter $\cp$ in \eqref{eq:cp} which is connected to the one in the
thrust distribution, see \eqref{eq:thrust}.  The leading power
correction is similar to the ones in EEC: the expected $1/Q$ behaviour
of the leading NP correction turns into $1/Q^{\gam}$ with $\gam$ a
slowly varying function of $Q$. We have $\gam\simeq 0.62$ for three quark
flavours, see \eqref{eq:gam}.

In conclusion, this paper is a first step towards the full QCD
analysis of azimuthal correlations in hard processes, which are among
the most interesting and least studied observables in QCD. Resummation
and NP corrections are both quite large. A detailed experimental
analysis (yet missing) would therefore be important in order to test
our knowledge of QCD dynamics and confinement effects in multi-jet
ensembles.

\section*{Acknowledgments}
We are grateful to Yuri Dokshitzer for helpful discussions and
suggestions, and to Giulia Zanderighi and Gavin Salam for support in
the numerical analysis.

\newpage
\appendix

\section{Elementary partonic cross-sections \label{App:El}}
For the elementary process \eqref{eq:El-proc} we may write the parton
momenta $P_1$, $P_2$ as
\begin{equation}
  \label{eq:parton}
\begin{split}
& P_1 = \frac{Q}{2x}(1,0,0,-1) \>,\quad
P_2 = \frac{Q}{2}(z_0,0,T_M,z_3) \>,\quad
\\& 
z_0 = \xi\!+\!\frac{(1\!-\!\xi)(1\!-\!x)}{x} \>,\quad
z_3 = \xi\!-\!\frac{(1\!-\!\xi)(1\!-\!x)}{x} \>,
\end{split}
\end{equation}
in terms of the variables in \eqref{eq:El-kin}. Distinguishing $P_2$
and $P_3$ according to \eqref{eq:P2P3}, the variable $\xi$ is given
in terms of $T_M$ by
\begin{equation}
\label{eq:psiT_M}
\xi = \frac{1}{2}\left[1-\sqrt{1-\frac{x T_M^2}{1-x}}\right]\>,
\end{equation}
and in terms of $y_2$ by 
\begin{equation}
\label{eq:ispxy3}
\begin{split}
&\xi=\frac{1-x}{2(2x-1)}\left[
\sqrt{1+\frac{4x(2x-1)y_2}{(1-x)^2}}-1\right]\>,\quad\left\{
\begin{split}
& 0<x<\frac{3}{4},\quad 0<y_2<\frac{1}{4x}\\
& \frac{3}{4}<x<1,\quad 0<y_2<\frac{2(1-x)^2}{x(2x-1)}
\end{split}\right.\\
&\xi=\frac{x^2 y_2-(1-x)^2}{(2x-1)(1-x+x y_2)}\>,\qquad
\qquad\qquad\qquad\> \frac{3}{4}<x<1,\>
\frac{2(1-x)^2}{x(2x-1)}<y_2<\frac{1-x}{x}
\end{split}
\end{equation}
The phase-space in terms of the variables $(x,\xi)$ then becomes:
\begin{equation}
\label{eq:isp_pm}
\begin{split}
&\xB<x<x_M\>,\quad x_M=\left\{\begin{split}
&\frac{1}{1+y_-}\>,\>\>\>y_-<1/3\\
&\frac{1}{4y_-}\>,\>\>\>\>\>\>\>\>y_->1/3
\end{split}\right.\>,\\
&\xi_-<\xi<\xi_+\>,\quad
\xi_-=\xi(x,y_-)\>,\quad
\xi_+=\left\{\begin{split}
&\xi(x,y_+)\>,\>\>\>x<{\textstyle\frac{1}{4y_+}}\\
&\half\>,\qquad\quad\>\>x>{\textstyle\frac{1}{4y_+}}
\end{split}\right. .
\end{split}
\end{equation}

We consider now the nature of the involved hard primary partons. We
identify the incoming parton of momentum $P_1$ by the index
$\can=q,\bar q,g$. Since \eqref{eq:P2P3} distinguishes $P_2$ from
$P_3$, in order to completely fix the configurations of the three
primary partons, we need to give an additional index $\conf=1,2,3$
identifying the gluon.  Therefore the primary partons with momenta
$\{P_1,P_2,P_3\}$ are in the following five configurations
\begin{equation}
\label{eq:configs}
\begin{split}
&\can=g,\,\conf=1\>\>\to\>\>\{gq\bar q\}\,,
\quad \mbox{or}\quad\{g\bar q q\}\,\\
&\can=q,\,\conf=2\>\>\to\>\>\{qgq\}\,, \\
&\can=\bar q,\,\conf=2\>\>\to\>\>\{\bar q g\bar q\}\,, \\
&\can=q,\,\conf=3\>\>\to\>\>\{qqg\}\,, \\
&\can=\bar q,\,\conf=3\>\>\to\>\>\{\bar q\bar q g\}\,.
\end{split}
\end{equation}

Next we give the corresponding five elementary distributions
$d\hat\sigma_{\can,\conf,f}$, with $f$ the fermion flavours.  
Since we consider only diagrams with the exchange of a single photon or
$Z_0$, the individual partonic cross-sections may be decomposed 
according to:
\begin{equation}
\label{eq:partonicxs}
\begin{split}
\frac{d\hat{\sigma}_{\can,\conf,f}}{dx\,d\xi\,dQ^2} = 
\frac{\al^2\as}{Q^4}
& \Bigl\{C^f(Q^2)\left[\left(2-2y+y^2\right)C_T^{\can,\conf}(x,\xi)
+2(1-y)C_L^{\can,\conf}(x,\xi)\right]\\ 
& +D^f(Q^2)y(2-y)C_3^{\can,\conf}(x,\xi)\Bigr\}\>,
\end{split}
\end{equation}
into transverse, longitudinal and parity-violating terms.
Here $y$ is defined by
\begin{equation}
\label{eq:y}
y = \frac{(Pq)}{(PP_e)} = \frac{Q^2}{\xB s} \>,
\end{equation}
where $P_e$ is the momentum of the incident electron, and
$s$ is the centre-of-mass energy squared of the collision (we neglect
the proton mass).
The flavour-dependent functions $C^f(Q^2)$ and $D^f(Q^2)$ show how the
$\gam$ and $Z$ exchange diagrams combine:
\begin{equation}
\label{eq:Z-cross}
\begin{split}
C^f(Q^2) &= e_f^2-\frac{2e_f V_f V_e}{\sin^2 2\theta_W}
\left(\frac{Q^2}{Q^2\!+\!M^2}\right)+\frac{(V_f^2\!+\!A_f^2)(V_e^2\!+\!A_e^2)}
{\sin^4 2\theta_W}\left(\frac{Q^2}{Q^2\!+\!M^2}\right)^2\>,\\
D^f(Q^2) &= -\frac{2e_f A_q A_e}{\sin^2 2\theta_W}
\left(\frac{Q^2}{Q^2\!+\!M^2}\right)+\frac{4V_f A_f V_e A_e}
{\sin^4 2\theta_W}\left(\frac{Q^2}{Q^2\!+\!M^2}\right)^2.
\end{split}
\end{equation}
The functions $C_{T/L/3}^{\can,\conf}(x,\xi)$ are the one loop 
QCD elementary square matrix elements for the hard vertex (there is no
contribution of order $\cO{\as^0}$ since we require transverse jets).
We have \cite{dishard}
\begin{equation}
\begin{split}
&C_T^{q,3} = C_F\left[\frac{x^2+\xi^2}{(1-x)(1-\xi)}
+2(1+x\xi)\right]\,,
\hspace{0.4cm} 
C_T^{g,1} = \left[x^2+(1-x)^2\right]
\frac{\xi^2+(1-\xi)^2}{\xi(1-\xi)}\,,\\
&C_L^{q,3} = C_F\cdot 4x\xi\,, 
\hspace{4.8cm} 
C_L^{g,1} = 8x(1-x)\,, \\
&C_3^{q,3} = C_F\left[\frac{x^2+\xi^2}{(1-x)(1-\xi)}
+2(x+\xi)\right]\,,
\hspace{0.5cm} 
C_3^{g,1} = 0\,,\\
&C_{T/L/3}^{q,2}(x,\xi) = C_{T/L/3}^{q,3}(x,1\!-\!\xi)\,,\>\>
C_{T/L}^{\bq,\conf}(x,\xi) = C_{T/L}^{q,\conf}(x,\xi)\,,\>\> 
C_3^{\bq,\conf}(x,\xi) = -C_3^{q,\conf}(x,\xi)\,.
\end{split}
\end{equation}

If we consider only photon exchange the index $\can$ is redundant, and
the relevant elementary distributions $d\hat{\sigma}_{\conf,f}^{(\gam)}$
are
\begin{equation}
\label{eq:dsigma-gamma1}
\begin{split}
&\frac{d\hat{\sigma}_{\conf,f}^{(\gam)}}{dx\,d\xi\,dQ^2}= 
\frac{\alpha^2\as}{Q^4}e_f^2
\left[\left(2\!-\!2y\!+\!y^2\right)\,C_T^{\conf}(x,\xi)
+2(1\!-\!y)\,C_L^{\conf}(x,\xi)\right]\>,\\
&C_T^{3}=C_F\left[\frac{x^2\!+\!\xi^2}{(1\!-\!x)(1\!-\!\xi)}
+2(1\!+\!x\xi)\right]\>,\quad
C_T^{1} =\left[x^2+(1\!-\!x)^2\right]
\frac{\xi^2+(1\!-\!\xi)^2}{\xi(1-\xi)}\>,\\
&C_L^{3}= C_F\cdot 4x\xi \>, \qquad C_L^{1} = 8x(1\!-\!x)\>,
\qquad C_{T/L}^{2}(x,\xi)\!=\!C_{T/L}^{3}(x,1\!-\!\xi)\>.
\end{split}
\end{equation}

\section{Observable decomposition\label{App:obs}}
Here we decompose the observable $H(\chi)$, defined in \eqref{eq:H},
in the case where we have two hard outgoing partons $p_2$ and $p_3$
accompanied by soft radiation $k_i$. Replacing the sum over hadrons in
\eqref{eq:H} by a sum over partons, and neglecting terms of 
$\cO{k_i^2/Q^2}$, we obtain
\begin{equation}
\begin{split}
H(\chi) =& \frac{p_{t2}^2+p_{t3}^2}{Q^2}\delta(\chi-\pi)
+2\frac{p_{t2}p_{t3}}{Q^2}\delta(\chi-\chi_{23})\\
&+2\sum_{i}\frac{p_{t2}k_{ti}}{Q^2}\delta(\chi-\chi_{2i})
+2\sum_{i}\frac{p_{t3}k_{ti}}{Q^2}\delta(\chi-\chi_{3i})\>,
\end{split}
\end{equation}
with $\chi_{ab}=\pi-|\phi_{ab}|$. 

In order to perform the calculation, we must introduce the definition
of the event plane as the plane formed by the incoming proton
direction $\vec{n}_P$ in the Breit frame and the unit vector
$\vec{n}_M$ which enters the definition of thrust major
\begin{equation}
  \label{eq:TM}
T_M = \max_{\vec{n}_M}\frac{1}{Q}{\sum_h}|\vec{p}_h\cdot\vec{n}_M|\>,
\qquad \vec{n}_M\cdot \vec{n}_P=0\>.
\end{equation}
For dijet events with transverse momentum $P_t\sim Q$ we have
$T_M=\cO{1}$. Fixing the event plane as the $yz$
plane, let the transverse momentum components of the outgoing partons
be
\begin{equation}
\vec{p}_{t2} = P_t (\eps_1, 1-\eps_2)\>, \quad
\vec{p}_{t3} = -P_t (\eps_3, 1-\eps_4)\>, \quad
\vec{k}_{ti} = (k_{xi},k_{yi}) = k_{ti}(\sin\phi_i,\cos\phi_i)\>,
\end{equation}
with $P_t = \half QT_M$.
Conservation of momentum gives
\begin{equation}
\phi_x = \eps_3-\eps_1 = \sum_i \frac{k_{xi}}{P_t}\>, \qquad
\eps_2-\eps_4 = \sum_i \frac{k_{yi}}{P_t}\>,
\end{equation}
while the definition \eqref{eq:TM} of $T_M$ gives
\begin{equation}
\eps_1+\eps_3 = -\sum_i \frac{k_{xi}}{P_t}
\left[\theta(k_{yi})-\theta(-k_{yi})\right]\>,\qquad
\eps_2+\eps_4 = \sum_i \frac{|k_{yi}|}{P_t} \>.
\end{equation}
Thus
\begin{equation}
\begin{split}
\eps_1 = -\sum_i \frac{k_{xi}}{P_t}\theta(k_{yi}) \qquad
&\eps_2 = \sum_i \frac{k_{yi}}{P_t}\theta(k_{yi}) \\
\eps_3 = \sum_i \frac{k_{xi}}{P_t}\theta(-k_{yi}) \qquad
&\eps_4 = -\sum_i \frac{k_{yi}}{P_t}\theta(-k_{yi}) \>.
\end{split}
\end{equation}

Expanding up to terms linear in the soft momenta, we get
\begin{equation}
\begin{split}
&\chi_{23} = |\eps_3-\eps_1| \qquad
\chi_{2i} = |\pi+\eps_1-\phi_i| \qquad
\chi_{3i} = |\phi_i-\eps_3| \\
&p_{t2} = P_t (1-\eps_2) \qquad
p_{t3} = P_t (1-\eps_4)
\end{split}
\end{equation}

Hence the observable $H(\chi)$ is given by
\begin{equation}
\begin{split}
H(\chi) = &\frac{2P_t^2}{Q^2} (1-\eps_2-\eps_4)\left[
\delta\left(\chi-|\eps_3-\eps_1|\right)+\delta(\chi-\pi)\right]\\
&+\frac{2P_t}{Q^2}\sum_i k_{ti}\left[\delta(\chi-|\pi+\eps_1-\phi_i|)+
\delta(\chi-|\phi_i-\eps_3|)\right]\>.
\end{split}
\end{equation}
The first term is generated by the two hard partons, but has
contributions both from the underlying hard momenta and from the soft
parton recoil. The second term is the direct contribution from soft
radiation.

Let us therefore take the recoil contribution from the first term (that
proportional to $\eps_2+\eps_4$) and add it to the soft term:
\begin{equation}
\begin{split}
H(\chi) &= \frac{2P_t^2}{Q^2}\left[\delta(\chi-|\eps_3-\eps_1|)
+\delta(\chi-\pi)\right]+\frac{2P_t}{Q^2}\sum_i k_{ti}\Bigl[
\delta(\chi-|\pi+\eps_1-\phi_i|)\\
&+\delta(\chi-|\phi_i-\eps_3|)-|\cos\phi_i|
\delta(\chi-|\eps_3-\eps_1|)-|\cos\phi_i|\delta(\chi-\pi)\Bigr]\>.
\end{split}
\end{equation}
Note the cancellation of collinear singularities: when $k_i$ is
collinear to $p_1$ we have $\phi_i=\eps_1$, which generates
cancellation between the second and third delta-functions in the soft
term above, and between the first and fourth. (We have $\cos\phi_i=1$
to this accuracy.)
Similarly, when $k_i$ is collinear to $p_2$ we have
$\phi_i=\pi+\eps_3$, which generates cancellation between the first
and third delta-functions, and between the second and fourth. (Now we
have $\cos\phi_i=-1$.)

The above is valid for all $\chi$ up to $\chi=\pi$, but we are only
interested in small $\chi$ for the resummation and leading power
correction, with larger values of $\chi$ given by matching with the
fixed order calculation. For small $\chi$, we may write
\begin{equation}
H(\chi) \simeq \frac{2P_t^2}{Q^2}\delta(\chi-|\phi_x|)
+\frac{2P_t}{Q^2}\sum_i k_{ti}\Bigl[\delta(\chi-|\bar{\phi}_i-\phi_x|)
-|\cos\bar{\phi}_i|\delta(\chi-|\phi_x|)\Bigr]\>,
\end{equation}
where instead of $\phi_i$ we now use $\bar{\phi}_i$, the azimuthal
angle between $k_i$ and either the hard parton $p_2$ or $p_3$:
\begin{equation}
\bar{\phi}_i=\left\{
\begin{split}
&\phi_{i2}=\phi_i-\eps_1 \qquad  \mbox{ for $k_i$ near $p_2$,} \\
&\phi_{3i}=\pi-\phi_i+\eps_3 \>  \mbox{ for $k_i$ near $p_3$.}
\end{split}
\right.
\end{equation}
The cancellation of the collinear singularity is still manifest as
we take $\bar\phi_i\to 0$.

\section{DL approximation\label{App:sud}}
Here we give an instructive simplified study of the behaviour of
the small and large-$b$ contributions to the distribution
$\cI(\chi)$. We use the DL radiation factor evaluated at 
$P_t=\frac{1}{2}Q$, with a fixed value of the running coupling ($\as=0.119$)
and without parton distribution functions:
\begin{equation}
\label{eq:DLrad}
\cF_{\rm DL}(b) = e^{-R_{\rm DL}(b)}\;,\qquad
R_{\rm DL}(b) = \frac{\as C_T}{\pi}\ln^2(\bar{b}P_t)\;,\qquad 
\bar b=b\> e^{\gam_E}\>.
\end{equation}
We separate the $b$-integral into a soft and hard contribution as
indicated in equation \eqref{eq:cIpm}, and discuss each contribution
in turn.

\subsection{Large $b$ (soft) contribution}
The large $b$ or soft contribution is that arising from multiple gluon
emissions with all $|k_x|<\chi P_t$, whose cumulative effect is a hard
parton recoil of $\chi P_t$. Thus we expect to see in this
contribution the usual Sudakov behaviour. We obtain
\begin{equation}
\begin{split}
\cI^{(+)}(\chi) &= \frac{d}{d\chi}
\int_{\frac{1}{\bar{\chi}P_t}}^{\infty}\frac{db}{\pi b}\sin(P_t b\chi)
\,e^{-R_{\rm DL}(b)}\\
&\approx \frac{d}{d\chi}\cdot e^{-R_{\rm DL}(\frac{1}{\bar{\chi}P_t})}
\left[\frac{e^{-\gamma_E R'}\sec\frac{\pi}{2}R'}{2\Gamma(1+R')}-
\frac{1}{\pi}\sum_{n=0}^\infty\frac{(-1)^n}{(2n+1)!}
\frac{e^{-(2n+1)\gamma_E}}{2n+1-R'}\right]\>,
\end{split}
\end{equation}
where the function $R'$ is
\begin{equation}
R' = \frac{2\as C_T}{\pi}\ln\frac{1}{\chi}\>,
\end{equation}
and the quantity in brackets is a SL function which equals
$\frac{1}{2}$ at DL level. The result is well-defined and convergent
for all $R'$, since the poles from the $\sec$ function are cancelled
explicitly by the second term in the brackets. (If however we attempt
the same procedure on the integral over the whole $b$ range, we obtain
only the first term, which diverges at $R'=1$.)

The difference between the approximation and the exact value of the
integral is formally beyond SL and is compensated for by the matching 
with fixed order. The Sudakov behaviour is clearly seen, along with the 
peak at $R'\approx 1$ (see figure \ref{fig:dl}).

\subsection{Small $b$ (hard) contribution}
The small $b$ or hard piece is that arising from two or more hard
gluon emissions, with $|k_x|>\chi P_t$, but which combine with
opposite signs to give the hard parton recoil of only $\chi P_t$. We
can expand in powers of $\chi$:
\begin{equation}
\begin{split}
\cI^{(-)}(\chi) &= \frac{d}{d\chi}\int_0^{\frac{1}{\bar{\chi}P_t}}
\frac{db}{\pi b}\sin(P_t b\chi)\,e^{-R_{\rm DL}(b)}\\
&= \frac{1}{\pi}\frac{d}{d\chi}\sum_{n=0}^\infty
\frac{(-1)^n(\chi P_t)^{2n+1}}{(2n+1)!}
\int_0^{\frac{1}{\bar{\chi}P_t}}db\,b^{2n}e^{-R_{\rm DL}(b)}\>,
\end{split}
\end{equation}
and then integrate to give
\begin{equation}
\begin{split}
\cI^{(-)}(\chi) = \frac{1}{\pi}\frac{d}{d\chi}\sum_{n=0}^\infty
\frac{(-1)^n(\chi e^{-\gam_E})^{2n+1}}{(2n+1)!}
\frac{e^{\frac{(2n+1)^2}{2a}}}{\sqrt{a}}
\Phi\left(\sqrt{a}\ln\frac{1}{\chi}-\frac{2n+1}{\sqrt{a}}\right)\>,
\end{split}
\end{equation}
where the quantity $a$ and the function $\Phi(x)$ are
\begin{equation}
a = \frac{2\as C_T}{\pi}\;,\qquad
\Phi(x) = \int_{-\infty}^x dt\,e^{-\frac{1}{2}t^2}\>.
\end{equation}
The series is rapidly convergent since as $n$ increases the argument
of the $\Phi$ function becomes more negative.  For this contribution
there is no Sudakov behaviour, but rather at $\chi=0$ the distribution
increases to a constant that behaves as $1/\sqrt{\as}$.

In figure \ref{fig:dl} are plotted the two contributions to $\cI$ as well
as the total. Notice that at very small $\chi$ (i.e. with $R'>1$) the soft 
radiation is suppressed and the distribution is due to hard emission. 
Soft radiation contributes around and above the Sudakov peak at $R'=1$.

\EPSFIGURE[ht]{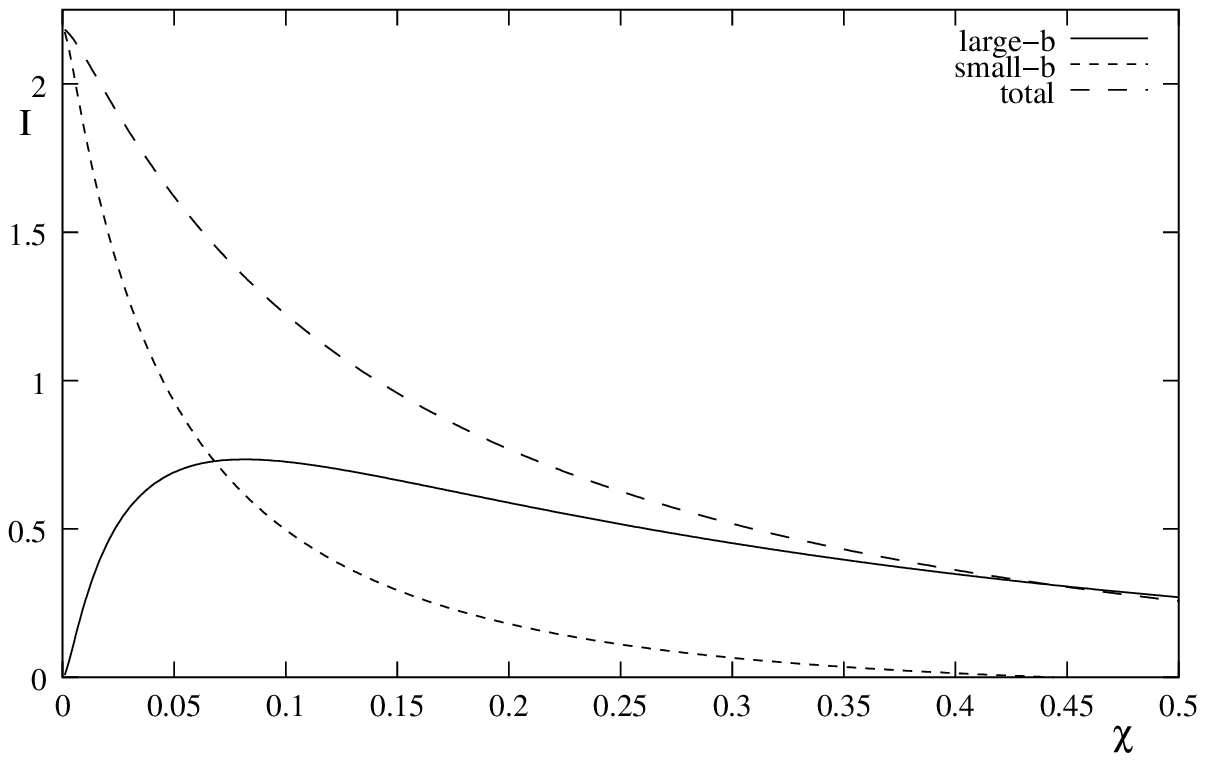}{Hard, soft and total contributions to $\cI(\chi)$
in the DL approximation for the azimuthal correlation.\label{fig:dl}}

\subsection{Comparison with EEC}
The case of energy-energy correlation in $\ee$ is similar to the
above, except that instead of $C_T$ the total colour charge of the
hard partons is $2 C_F$, and the integration measure is substituted
according to
\begin{equation}
\frac{db}{\pi b}\sin(P_t b \chi)\longrightarrow
Q\chi db \,J_1\left(\frac{Q}{2}b\chi\right)\;.
\end{equation}
In particular the 2-dimensional phase-space ensures that the distribution
peaks and falls linearly to zero, while the fact that the SL approximation
is valid up to $R'=2$ means that it is a good approximation for in the 
region of the peak, up to matching with fixed order.

In figure \ref{fig:dl2} are plotted large and small-b contributions
and the total for this distribution. The large-b contribution
dominates until well below the peak. This very much simplified study
shows the qualitative behaviour of the distribution in $b$-space, and
gives a helpful insight into the physics of the azimuthal correlation.

\EPSFIGURE[ht]{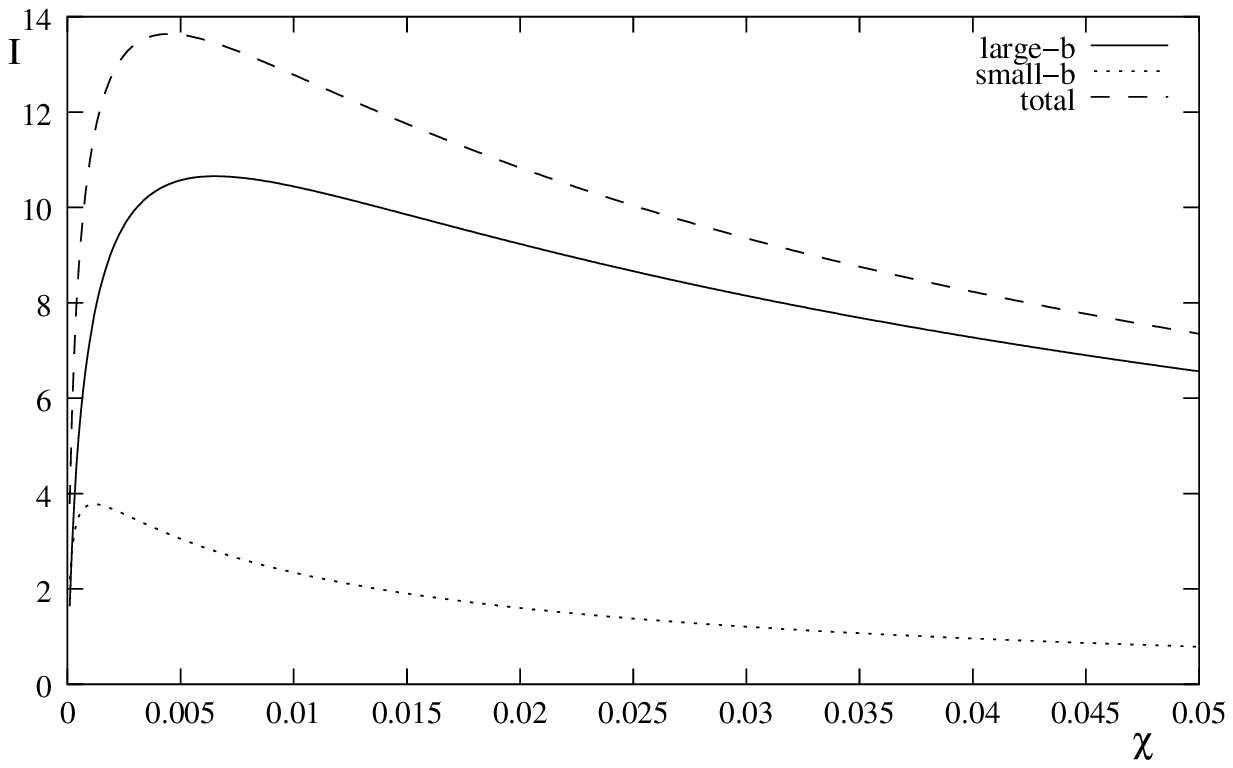}{Hard, soft and total contributions to
  $\cI(\chi)$ in the DL approximation for the EEC
  distribution.\label{fig:dl2}}

\section{Soft contribution \label{App:Bsoft}}
In this appendix we study the perturbative and non-perturbative
contributions to the quantity $B_{\conf}(b)$ given in \eqref{eq:B}.
We consider each dipole separately.

\subsection{Dipoles 12 and 13}
Consider
\begin{equation}
B_{12}(b) = \int\frac{d^3 k}{\pi\omega}w_{12}(k)\,\frac{k_t}{P_t}
\left[\cos(P_t b\bar{\phi})-|\cos\bar{\phi}|\right]\>.
\end{equation}
Denoting by $p^*_1,p^*_2$ and $k^*$ the momenta in this system, 
we introduce the Sudakov decomposition
\begin{equation}
\label{eq:Sud}
p_2^*=\frac{Q_{ab}}{2}(1,0,0,1)\>,\quad 
p_1^*=\frac{Q_{ab}}{2}(1,0,0,-1)\>,\qquad 
k^*=\al p_2^*+\be p_1^*+\ka\>, 
\end{equation}
where the scale $Q_{12}^2=2(p_1p_2)$.  Here the two-dimensional vector
$\vka=(\ka_x,\ka_y)$ is the transverse momentum orthogonal to the
$12$-dipole momenta ($\ka^2=k^2_{12,t}$). We have then
\begin{equation}
\label{eq:12-cm}
w_{12}(k)=\frac{\as(\ka^2)}{\pi \ka^2}\>, \qquad
\frac{d^3k}{\pi\omega}=\frac{d^2\ka}{\pi}\frac{d\al}{\al}\>, \quad 
\al>\frac{\ka^2}{Q^2_{12}}\>.
\end{equation}

We must also express the Breit-frame components of $k$ in terms of these
variables. The Sudakov variables $\al$ and $\be$ are Lorentz invariant,
while the two-component vector $\vka$ transforms into a 4-vector
with components
\begin{equation}
\ka_0^{\mbox{\scriptsize Breit}}=\frac{P_t}{Q\xi}\ka_y\>,\quad
\ka_x^{\mbox{\scriptsize Breit}}=\ka_x\>,\quad
\ka_y^{\mbox{\scriptsize Breit}}=\ka_y\>,\quad
\ka_z^{\mbox{\scriptsize Breit}}=-\frac{P_t}{Q\xi}\ka_y\>.
\end{equation}
Thus the transverse momentum components of $k$ in the Breit frame are
\begin{equation}
k_x = \ka_x \>,\qquad
k_y = \al P_t + \ka_y \>.
\end{equation}
We take $\bar\phi$ as the azimuthal angle between $p_2$ and $k$ ---
this neglects the contribution from $k$ near to $p_3$, which
contributes beyond SL and so is included by matching with the fixed
order result. Therefore
\begin{equation}
\cot\bar\phi = \frac{k_y}{k_x} = \frac{\ka_y}{\ka_x}+\frac{\al P_t}{\ka_x}
=\cot\phi + \frac{\al P_t}{\ka}\csc\phi \>,
\end{equation}
and the $12$-dipole contribution is
\begin{equation}
B_{12}(b) = \int\frac{d\ka^2}{\ka^2}\frac{d\phi}{2\pi}
\frac{d\al}{\al}\frac{\as(\ka)}{\pi}\cdot\frac{\ka\sin\phi}{P_t\sin\bar\phi}
\left[\cos(P_t b\bar{\phi})-|\cos\bar{\phi}|\right]\>.
\end{equation}

Changing integration variable from $\al$ to $\bar\phi$ gives
\begin{equation}
\label{eq:B12}
\begin{split}
&B_{12}(b) = \frac{2}{\pi P_t}\int d\ka\,\as(\ka)\int_0^\pi
\frac{d\phi}{\pi}\sin^2\phi\int_{\bar\phi_m}^{\bar\phi_M}
\frac{d\bar\phi}{\sin(\phi\!-\!\bar\phi)}
\cdot\frac{\cos(P_t b\bar{\phi})-|\cos\bar{\phi}|}{\sin^2\bar\phi}\>,\\
&\cot\bar\phi_m = \cot\phi + \frac{P_t}{\ka}\csc\phi \>,\qquad
\cot\bar\phi_M = \cot\phi + \frac{P_t\ka}{Q_{12}^2}\csc\phi \>,
\end{split}
\end{equation}
which is clearly beyond SL, since both the soft logarithm ($\ka\to0$)
and the collinear logarithm ($\bar\phi\to0$) are cancelled.
The $13$-dipole is treated similarly.

\subsection{Dipole 23}
The $23$-dipole is analysed along the same lines as the $12$, except that
there are some minor differences.
Again we introduce the Sudakov decomposition
\begin{equation}
\label{eq:Sud23}
p_2^*=\frac{Q_{23}}{2}(1,0,0,1)\>,\quad 
p_3^*=\frac{Q_{23}}{2}(1,0,0,-1)\>,\qquad 
k^*=\al p_2^*+\be p_3^*+\ka\>.
\end{equation}
We must again express the Breit-frame components of $k$ in terms of these
variables, but now the two-component vector $\vka$ transforms into a 4-vector
with components
\begin{equation}
\ka_0^{\mbox{\scriptsize Breit}}=
\frac{P_t(2x\!-\!1)}{Q(1\!-\!x)}\ka_y\>,\quad
\ka_x^{\mbox{\scriptsize Breit}}=\ka_x\>,\quad
\ka_y^{\mbox{\scriptsize Breit}}=(1\!-\!2\xi)\ka_y\>,\quad
\ka_z^{\mbox{\scriptsize Breit}}=\frac{P_t}{Q(1\!-\!x)}\ka_y\>.
\end{equation}
The transverse momentum components of $k$ in the Breit frame are then
\begin{equation}
k_x = \ka_x \>,\qquad
k_y = (\al-\be) P_t + (1-2\xi)\ka_y \>.
\end{equation}

We take $\bar\phi$ as the azimuthal angle between $p_2$ and $k$ in the
region where $\al>\be$, and as the angle between $p_3$ and $k$ when
$\al<\be$. The neglected contributions are beyond SL and so are
included by matching with the fixed order result. We take each
hemisphere separately: in the right hemisphere ($\al>\be$)
\begin{equation}
\cot\bar\phi = (1-2\xi)\cot\phi + \frac{(\al-\be)P_t}{\ka}\csc\phi \>,
\end{equation}
and the $23$-dipole right hemisphere contribution is
\begin{equation}
B_{23}^R(b) = \int_{\al>\be}\frac{d\ka^2}{\ka^2}\frac{d\phi}{2\pi}
\frac{d\al}{\al}\frac{\as(\ka)}{\pi}\cdot\frac{\ka\sin\phi}{P_t\sin\bar\phi}
\left[\cos(P_t b\bar{\phi})-|\cos\bar{\phi}|\right]\>.
\end{equation}

Changing integration variable from $\al$ to $\bar\phi$ gives
\begin{equation}
\label{eq:B23R}
\begin{split}
&B_{23}^R(b) = \frac{2}{\pi P_t}\int d\ka\,\as(\ka)\int_0^\pi
\frac{d\phi}{\pi}\sin^2\phi\int_{\bar\phi_m}^{\bar\phi_M}
\frac{d\bar\phi}{\sqrt{f(\phi,\bar\phi)}}
\cdot\frac{\cos(P_t b\bar{\phi})-|\cos\bar{\phi}|}{\sin^2\bar\phi}\>,\\
&f(\phi,\bar\phi) = \left[\sin\phi\cos\bar\phi-(1\!-\!2\xi)
\cos\phi\sin\bar\phi\right]^2+4\frac{P_t^2}{Q_{23}^2}\sin^2\bar\phi\\
&\cot\bar\phi_m = (1-2\xi)\cot\phi + \frac{P_t}{\ka}\csc\phi \>,\qquad
\cot\bar\phi_M = (1-2\xi)\cot\phi \>,
\end{split}
\end{equation}
which is clearly beyond SL, since both the soft logarithm ($\ka\to0$)
and the collinear logarithm ($\bar\phi\to0$) are cancelled.
The left hemisphere is treated similarly.

\subsection{NP contribution}
We calculate the NP contribution using the standard procedure, see
\cite{DMW}.

Applying this to equation \eqref{eq:B12} gives the NP contribution to
$B_{12}$ as
\begin{equation}
\begin{split}
\delta B_{12}(b) =& \frac{\cM}{\pi}\int dm^2\,\daef(m^2)\frac{-d}{dm^2}
\int_0^\infty\frac{d\ka^2}{P_t\sqrt{\ka^2+m^2}}
\int_0^\pi\frac{d\phi}{\pi}\sin^2\phi\\
&\times\int_{\bar\phi_m}^{\bar\phi_M}\frac{d\bar\phi}{\sin(\phi\!-\!\bar\phi)}
\cdot\frac{\cos(P_t b\bar{\phi})-|\cos\bar{\phi}|}{\sin^2\bar\phi}\>,\\
\cot\bar\phi_m =& \cot\phi + \frac{P_t}{\sqrt{\ka^2+m^2}}\csc\phi \>,\qquad
\cot\bar\phi_M = \cot\phi + \frac{P_t\sqrt{\ka^2+m^2}}{Q_{12}^2}\csc\phi \>.
\end{split}
\end{equation}
Here $\cM$ is the Milan factor \cite{milan} which accounts for the non
fully inclusive nature of the observable.

Doing the $m$-derivative
and $\ka$-integral gives
\begin{equation}
\begin{split}
\delta B_{12}(b) =& \frac{2\cM}{\pi P_t}\int dm\,\daef(m^2)
\int_0^\pi\frac{d\phi}{\pi}\sin^2\phi
\int_{\bar\phi_m}^{\bar\phi_M}\!\!\!\!
\frac{d\bar\phi}{\sin(\phi\!-\!\bar\phi)}
\cdot\frac{\cos(P_t b\bar{\phi})-|\cos\bar{\phi}|}{\sin^2\bar\phi}\>,\\
\cot\bar\phi_m =& \cot\phi + \frac{P_t}{m}\csc\phi \>,\qquad
\cot\bar\phi_M = \cot\phi + \frac{P_t m}{Q_{12}^2}\csc\phi \>.
\end{split}
\end{equation}

We wish to find the leading term in the expansion of this integral
in the range $m\ll b^{-1}\ll P_t$. Writing $\bar\phi = u/(bP_t)$ and
expanding the integrand yields
\begin{equation}
\begin{split}
\delta B_{12}(b) &= \frac{2\cM}{\pi P_t}\int dm\,\daef(m^2)
\int_0^\pi\frac{d\phi}{\pi}\sin\phi
\int_{0}^{\infty} bP_t du\frac{\cos u-1}{u^2}+\cdots\\
&= -b\cdot\cp+\cdots \qquad
\cp=\frac{2\cM}{\pi}\int dm\,\daef(m^2)\>,
\end{split}
\end{equation}
where the dots involves terms which are of higher order in $b$. 
The $13$-dipole gives an identical result.

The same procedure applied to equation \eqref{eq:B23R} gives the NP 
contribution to $B_{23}^R$ as
\begin{equation}
\begin{split}
\delta B_{23}^R(b) =& \frac{\cM}{\pi}\int dm^2\,\daef(m^2)\frac{-d}{dm^2}
\int_0^\infty\frac{d\ka^2}{P_t\sqrt{\ka^2+m^2}}
\int_0^\pi\frac{d\phi}{\pi}\sin^2\phi\\
&\times\int_{\bar\phi_m}^{\bar\phi_M}\frac{d\bar\phi}{\sqrt{f(\phi,\bar\phi)}}
\cdot\frac{\cos(P_t b\bar{\phi})-|\cos\bar{\phi}|}{\sin^2\bar\phi}\>,\\
\cot\bar\phi_m =& (1-2\xi)\cot\phi + 
\frac{P_t}{\sqrt{\ka^2+m^2}}\csc\phi \>,\qquad
\cot\bar\phi_M = (1-2\xi)\cot\phi \>.
\end{split}
\end{equation}

Doing the $m$-derivative and $\ka$-integral gives
\begin{equation}
\begin{split}
\delta B_{23}^R(b) =& \frac{2\cM}{\pi P_t}\int dm\,\daef(m^2)
\int_0^\pi\frac{d\phi}{\pi}\sin^2\phi
\int_{\bar\phi_m}^{\bar\phi_M}\!\!\!\!
\frac{d\bar\phi}{\sqrt{f(\phi,\bar\phi)}}
\cdot\frac{\cos(P_t b\bar{\phi})-|\cos\bar{\phi}|}{\sin^2\bar\phi}\>,\\
\cot\bar\phi_m =& (1-2\xi)\cot\phi + \frac{P_t}{m}\csc\phi \>,\qquad
\cot\bar\phi_M = (1-2\xi)\cot\phi \>.
\end{split}
\end{equation}

We wish to find the leading term in the expansion of this integral
in the range $m\ll b^{-1}\ll P_t$. Writing $\bar\phi = u/(bP_t)$ and
expanding the integrand yields
\begin{equation}
\begin{split}
\delta B_{23}^R(b) &= \frac{2\cM}{\pi P_t}\int dm\,\daef(m^2)
\int_0^\pi\frac{d\phi}{\pi}\sin\phi
\int_{0}^{\infty} bP_t du\frac{\cos u-1}{u^2}+\cdots\\
&= -b\cdot\cp +\cdots
\end{split}
\end{equation}
where, as before, the terms in the dots are of order $b^2$.
The left hemisphere gives an identical result, and so the total is
\begin{equation}
\delta B_{23}(b) = -2b\cdot\cp+\cdots
\end{equation}
Combining the three dipoles then gives the leading non-perturbative
contribution
\begin{equation}
\delta B_{\conf}(b) = -\left(C_2^{(\conf)}+C_3^{(\conf)}\right)b\cdot
\cp\>.
\end{equation}

The complete treatment of this correction requires the definition of
the coupling in the infra-red region (this will be done by the
dispersive method \cite{DMW}), the analysis of contributions which are
non fully inclusive, leading to the Milan factor \cite{milan}, and of
the merging PT and NP contributions to the observable in a renormalon
free manner. This can be done in the usual way and the NP parameter is
given by
\begin{equation}
  \label{eq:cp}
 \cp \!\equiv  \cM \frac{4}{\pi^2}\mu_I
\left\{ \alpha_0(\mu_I)- \bar{\alpha}_s
  -\beta_0\frac{\bar{\alpha}_s^2}{2\pi}\left(\ln\frac{Q}{\mu_I} 
+\frac{K}{\be_0}+1\right) \right\},  
\end{equation}
where 
\begin{equation}
\label{eq:K-beta0}
\bar{\alpha}_s\equiv \al_{\MSbar}(Q)\>,\quad
   K\equiv
  C_A\left(\frac{67}{18}-\frac{\pi^2}{6}\right)-\frac{5}{9}n_f \>,
  \quad \beta_0=\frac{11N_c}{3}-\frac{2n_f}{3}\>.
\end{equation}
The $K$ factor accounts for the mismatch between the $\MSbar$ and the
physical scheme \cite{CMW} and $\al_0(\mu_I)$ is the integral of the
running coupling over the infra-red region, see \cite{DMW}.  We have
that, at two loops, $\cp$ is independent of $\mu_I$. Assuming
universality of the coupling~\cite{DMW,SZ} this NP parameter is the same
as already measured in $2$-jet observables.  For instance, the shift
for the $\tau=1\!-\!T$ distribution is
\begin{equation}
  \label{eq:thrust}
  \frac{d\sigma}{d\tau}(\tau)=
\frac{d\sigma^{\PT}}{d\tau}(\tau\!-\!\Delta_{\tau})\>,
\qquad   \Delta_{\tau}=C_F\,c_{\tau}\cp\>,
\quad c_{\tau}=2\>,
\end{equation}
where $C_F$ enters due to the fact the $2$-jet system is made of
a quark-antiquark pair.

\section{Estimates of $\VEV{b}$ using 1-loop coupling \label{App:gam}}
We consider using the 1-loop running coupling
\begin{equation}
\label{eq:1-loop}
\as(k) = \frac{2\pi}{\be_0}\,\frac{1}{\ln(k/\Lambda)}\>.
\end{equation}
The DL function $r(\bar{b},Q)$ defined in \eqref{eq:PT-rad} then becomes
\begin{equation}
r(\bar{b},Q) = \frac{4}{\be_0}\left[\ln(Q/\Lambda)
\ln\frac{\ln(2Q/\Lambda)}{\ln(2/\bar{b}\Lambda)}-\ln(\bar{b}Q)\right]\>,
\qquad\bar{b}<\frac{2}{\Lambda}\>,
\end{equation}
and we impose $r(\bar{b},Q)=\infty$ for $\bar{b}>\frac{2}{\Lambda}$.
We also use parton density functions determined by the 1-loop
evolution equation
\begin{equation}
\frac{\partial\cP_N(k)}{\partial\ln k}=\frac{\as(k)}{\pi}
\gamma_N\cdot\cP_N(k)
\end{equation}
with $\gamma_N$ the leading-order anomalous dimension matrix, and
$\cP_N(k)$ the moments of the parton density functions at scale $k$.
This equation has solution
\begin{equation}
\cP_N(k) = \left(\frac{\ln(k/\Lambda)}{\ln(Q/\Lambda)}\right)^{
\frac{2}{\be_0}\gam_N}\cdot\cP_N(Q)
\end{equation}
We define the integrals
\begin{equation}
I_n^{(\conf)}(N) = 
\int_0^\infty db\,b^n\cP_N(2\bar{b}^{-1})e^{-R_{\conf}(\bar{b})}\>,
\qquad\bar{b}=be^{\gam_E}\>,
\end{equation}
using the radiator given in \eqref{eq:PT-rad}. The choice of 
$2\bar{b}^{-1}$ rather than $\bar{b}^{-1}$ as the factorisation scale
is merely for convenience --- the difference is beyond SL.
Evaluation yields
\begin{equation}
I_n^{(\conf)}(N) = \left(\frac{2e^{-\gam_E}}{\Lambda}\right)^n
\left(\frac{1+\frac{4}{\be_0}C_T}{n+1+\frac{4}{\be_0}C_T}\right)^
{1+\frac{2}{\be_0}\gam_N+\frac{4}{\be_0}\sum_{a=1}^3 C_a^{(\conf)}
\ln\frac{\zeta_a^{(\conf)}Q_a^{(\conf)}}{\Lambda}}\cdot I_0^{(\conf)}(N)
\end{equation}
with 
\begin{equation}
\begin{split}
I_0^{(\conf)} =& \frac{2e^{-\gam_E}}{\Lambda}
\left(\ln\frac{Q}{\Lambda}\right)^{-\frac{2}{\be_0}\gam_N}
\frac{\Gamma\left(1+\frac{2}{\be_0}\gam_N+\frac{4}{\be_0}\sum_{a=1}^3 
C_a^{(\conf)}\ln\frac{\zeta_a^{(\conf)}Q_a^{(\conf)}}{\Lambda}\right)}
{\left(1+\frac{4}{\be_0}C_T\right)^{1+\frac{2}{\be_0}\gam_N+
\frac{4}{\be_0}\sum_{a=1}^3 C_a^{(\conf)}
\ln\frac{\zeta_a^{(\conf)}Q_a^{(\conf)}}{\Lambda}}}\\
& \times\prod_{a=1}^3
\left(\frac{2\zeta_a^{(\conf)}Q_a^{(\conf)}}{\Lambda}\right)^
{\frac{4}{\be_0}C_a^{(\conf)}}\prod_{a=1}^3
\left(\ln\frac{2\zeta_a^{(\conf)}Q_a^{(\conf)}}{\Lambda}\right)^
{-\frac{4}{\be_0}C_a^{(\conf)}
\ln\frac{\zeta_a^{(\conf)}Q_a^{(\conf)}}{\Lambda}}
\cdot\cP_N(Q)
\end{split}
\end{equation}
The average value of $b$ is therefore, at fixed $N$, 
\begin{equation}
\langle b\rangle_N = \frac{I_1^{(\conf)}(N)}{I_0^{(\conf)}(N)}
\sim \frac{a_{\conf}(N)}{\Lambda}
\left(\frac{\Lambda}{Q}\right)^{\frac{4}{\be_0}C_T
\ln\frac{2+\frac{4}{\be_0}C_T}{1+\frac{4}{\be_0}C_T}}
\end{equation}
with a coefficient $a_\conf$ that depends on $N$. The exponent is
$N$-independent and then we obtain the DL estimate in \eqref{eq:gam}.

\section{Formul\ae\ for numerical analysis\label{App:Matc}}

In this appendix we report some analytical formulae needed to perform
the numerical analysis described in section~\ref{sec:num}.

Within SL accuracy the PT radiator \eqref{eq:PT-rad} can be written as
\begin{equation}
\label{eq:r-SL}
R_{\conf}(b)= C_T \>r_1(\bar b)-C_T \ln 2 \>r'(\bar b)+
\sum_a C^{(\conf)}_a \ln \frac{Q^{(\conf)}_a\zeta^{(\conf)}_a}{Q} \>
r_2(\bar b)\>, \quad\bar b=b e^{\gam_E}\>, 
\end{equation}
where $Q_a^{(\conf)}$, $\zeta_a^{(\conf)}$ and $C_a^{(\conf)}$ are
introduced in \eqref{eq:QCs}, while $r_1,r_2$ and $r'$ are given by:
\begin{equation} 
\label{eq:rad-calc}
\begin{split}
r_1(\bar b)&=\int_{1/\bar b}^Q \frac{dk}{k}
\frac{2\as(k)}{\pi}\ln\frac{Q}{k}=
-\frac{8\pi}{\be_0^2\bar{\as}}\left(\ell+\ln(1-\ell)\right)+\frac{4K}{\be_0^2}
\left(\ln(1-\ell)+\frac{\ell}{1-\ell}\right)\\
\\&-\frac{4\be_1}{\be_0^3}\left(\frac12\ln^2(1-\ell)+
\frac{\ln(1-\ell)+\ell}{1-\ell}\right)\>,\qquad
\ell=\frac{\bar{\as}}{2\pi}\be_0\ln\bar bQ\>,\\
r_2(\bar b)&=\int_{1/\bar b}^Q \frac{dk}{k}\frac{2\as(k)}{\pi}=
-\frac{4}{\be_0}\ln(1-\ell)\>,\qquad
r'(\bar b)=\frac{2\as(\bar b^{-1})}{\pi}\ln\bar bQ =\frac{4}{\be_0}\frac{\ell}{1-\ell}\>.
\end{split}
\end{equation}
In the above expressions $\be_0$ can be found in \eqref{eq:K-beta0}
and $\be_1$ is given by:
\begin{equation}
  \label{eq:betas}
  \be_1 = \frac{17 C_A^2 -5 C_A n_f - 3 C_F n_f}{3}\>.
\end{equation}

The coupling $\bar\as$ is in the $\MSbar$ scheme
and the constant $K$ relating the physical scheme \cite{CMW} to the
$\MSbar$ is defined in \eqref{eq:K-beta0}.

Actually \eqref{eq:rad-calc} makes sense only for $\ell<1$, so that,
as usual (see \cite{EEC}), we impose
\begin{equation}
e^{-R(b)}=0\>,  \quad \mbox{for} \>\>\bar b Q>\exp\left(\frac{2\pi}{\bar{\as}\be_0}\right) \>.
\end{equation}

We report also the expression for the coefficients $G_{11}$ and
$c_1^{\res}$ introduced in \eqref{eq:Sigma-ev-first}:
\begin{equation}
\label{eq:g11-c1res}
\begin{split}
G_{11}(y_{\pm})&= -\sigma^{-1}(y_{\pm})
\sum_{\rho}\int_{\xB}^{x_M}\frac{dx}{x}\int_{\xi_-}^{\xi_+}d\xi
\left(\frac{d\hat\sigma_{\rho}}{dx d\xi dQ^2}\right)
\frac{4P_t^2}{Q^2}\cP_{\rho}\left(\frac{\xB}{x},Q\right)\\
&\cdot\left( 4\sum_a C^{(\conf)}_a\ln\frac{Q^{(\conf)}_a\zeta^{\conf}_a}{2 P_t}
+\frac{2\pi}{\as}
\frac{\partial\ln\cP_{\rho}}{\partial\ln Q}\right)\>,
\\
c_1^{\res}(y_{\pm})&=\sigma^{-1}(y_{\pm})
\sum_{\rho}\int_{\xB}^{x_M}\frac{dx}{x}\int_{\xi_-}^{\xi_+}d\xi
\left(\frac{d\hat\sigma_{\rho}}{dx d\xi dQ^2}\right)
\frac{4P_t^2}{Q^2}\cP_{\rho}\left(\frac{\xB}{x},Q\right)\\
&\cdot\left(\ln\frac{P_t}{Q}
\left( 4\sum_a C^{(\conf)}_a\ln\frac{Q^{(\conf)}_a\zeta^{\conf}_a}{2
Q}
+\frac{2\pi}{\as}
\frac{\partial\ln\cP_{\rho}}{\partial\ln Q}\right)
-C_T \frac{\pi^2}{6}-2 C_T \ln^2\frac{P_t}{Q}\right)\>.
\end{split}
\end{equation}

\end{document}